\documentclass{aastex62}

\usepackage{CJKutf8}

\usepackage{float}
\usepackage{upgreek}
\usepackage[shortlabels]{enumitem}
\usepackage[caption=false]{subfig}
\usepackage{graphbox}

\graphicspath{{./}{figures/}}

\received{}
\revised{}
\accepted{} 

\submitjournal{ApJS}

\shorttitle{Accuracy limits of spectral slope measurements in asteroid spectroscopy}
\shortauthors{Marsset et al.}

\begin{document}

\title{Twenty years of SpeX: Accuracy limits of spectral slope measurements in asteroid spectroscopy}

\correspondingauthor{Micha\"el Marsset}
\email{mmarsset@mit.edu}

\author[0000-0001-8617-2425]{Micha\"el Marsset}
\affil{Department of Earth, Atmospheric and Planetary Sciences, MIT, 77 Massachusetts Avenue, Cambridge, MA 02139, USA}

\author[0000-0002-8397-4219]{Francesca E. DeMeo}
\affiliation{Department of Earth, Atmospheric and Planetary Sciences, MIT, 77 Massachusetts Avenue, Cambridge, MA 02139, USA}
       
\author[0000-0002-9995-7341]{Richard P. Binzel}
\affiliation{Department of Earth, Atmospheric and Planetary Sciences, MIT, 77 Massachusetts Avenue, Cambridge, MA 02139, USA}

\author[0000-0003-4191-6536]{Schelte J. Bus}
\affiliation{Institute for Astronomy, University of Hawaii, 2860 Woodlawn Drive, Honolulu, HI 96822-1839, USA}

\author[0000-0001-8889-8692]{Thomas H. Burbine}
\affiliation{Department of Astronomy, Mount Holyoke College, South Hadley MA 01075}

\author[0000-0002-6423-0716]{Brian Burt}
\affil{Lowell Observatory, 1400 W. Mars Hill Road, Flagstaff, AZ, 86001, USA}

\author[0000-0001-6765-6336]{Nicholas Moskovitz}
\affil{Lowell Observatory, 1400 W. Mars Hill Road, Flagstaff, AZ, 86001, USA}

\author[0000-0002-6977-3146]{David Polishook}
\affiliation{Faculty of Physics, Weizmann Institute of Science, Israel}


\author[0000-0002-9939-9976]{Andrew S. Rivkin}
\affiliation{Johns Hopkins University Applied Physics Laboratory, Laurel, MD USA}

\author[0000-0003-3291-8708]{Stephen M. Slivan}
\affiliation{Department of Earth, Atmospheric and Planetary Sciences, MIT, 77 Massachusetts Avenue, Cambridge, MA 02139, USA}

\author[0000-0003-3091-5757]{Cristina Thomas}
\affiliation{Department of Astronomy and Planetary Science, Northern Arizona University, PO Box 6010, Flagstaff, AZ 86011.}

\begin{abstract}

We examined two decades of SpeX/NASA Infrared Telescope Facility observations from the Small Main-Belt Asteroid Spectroscopic Survey (SMASS) and the MIT-Hawaii Near-Earth Object Spectroscopic Survey (MITHNEOS) to investigate uncertainties and systematic errors in reflectance spectral slope measurements of asteroids. From 628 spectra of 11 solar analogs used for calibration of the asteroid spectra, we derived an uncertainty of $\upsigma_{s'}\,=\,4.2\%\,\upmu {\rm m}^{-1}$ on slope measurements over 0.8 to 2.4\,$\upmu$m. Air mass contributes to -0.92\%\,$\upmu$m$^{-1}$ per 0.1 unit air mass difference between the asteroid and the solar analog, and therefore for an overall $\,2.8\%\,\upmu {\rm m}^{-1}$ slope variability in SMASS and MITHNEOS designed to operate within 1.0--1.3 air mass. 
No additional observing conditions (including parallactic angle, seeing and humidity) were found to contribute systematically to slope change. 
We discuss implications for asteroid taxonomic classification works. 
Uncertainties provided in this study should be accounted for in future compositional investigation of small bodies to distinguish intrinsic heterogeneities from possible instrumental effects.

\end{abstract}

\section{Introduction}


Telescopic characterization of asteroids, the leftovers of planetary formation in our solar system, is key for advancing scientific knowledge about the origin and evolution of our solar system. 
To date, information about the compositional distribution of these bodies in the main belt and near-Earth space has mostly come from broad-band color photometry \citep{Chapman:1975, Gradie:1982, Gradie:1989, DeMeo:2013} and reflectance spectroscopy \citep{Bus:2002, Mothe-Diniz:2003, Lazzaro:2004, Binzel:2004, binzel:2019}, which consists of measuring the wavelength-dependent ratio between the incident and reflected light of the sun on the surface of an object. 
In the visible and near-infrared (NIR) wavelengths, spectroscopy provides a powerful diagnostic tool for the presence of silicate absorption bands near 1 and 2~$\upmu$m \citep{Gaffey:1993, Chapman1996, Binzel:2004}, hydration features between 0.6 and 0.9~$\upmu$m \citep{Vilas:1996, Fornasier:2014, Vernazza:2016}, and overall spectral slope (defined as the linear change in reflectance per unit wavelength) that can be diagnostic of surface composition and age (e.g., \citealt{Clark:2002, Brunetto:2015} and references therein). 
In return, such measurements allow distinguishing taxonomic classes of asteroids (e.g., \citealt{Bus:2002}) and performing quantitative study of their composition, e.g., by directly comparing measurements with laboratory data of meteorites \citep{Chapman1996, Burbine:2001} and/or synthetic spectra generated by use of a radiative transfer code \citep{Hapke:1993tt, Shkuratov:1999gpa}. 

It is well known, however, that a number of observing parameters can affect the throughput of a telescope, thereby introducing measurement artifacts that can be incorrectly interpreted as true properties of the targeted objects. 
Spectral slope measurements in particular are highly sensitive to instrumental and weather conditions, 
which adds difficulty to the classification and compositional interpretation for many classes of asteroids. 
Most notably affected are spectrally weakly-featured bodies, such as carbonaceous B-/C-types (e.g., Bennu, Ryugu), iron and stony-iron asteroids (e.g., Psyche), whose characterization in the visible and NIR mostly relies on spectral slope. 
While asteroid surveys are usually designed to mitigate these effects, e.g., by restricting observations to low air mass values and reasonable weather conditions, systematics in asteroid survey measurements can hardly be entirely removed. 
In this paper, we take advantage of two decades of observations acquired with the SpeX spectrograph \citep{Rayner:2003} on the Infrared Telescope Facility (IRTF) to explore the systematics and sources of errors in spectral slope measurements of asteroids, and evaluate consequences for asteroid taxonomy and compositional interpretation of individual objects. 
First, we present a brief history of the SpeX instrument in the context of the Solar System exploration (Section~\ref{sec:spexscience}). 
Then, we derive reliable uncertainties for spectral slope measurements of asteroids, and investigate systematic biases in those measurements (Section~\ref{sec:errors}). 
Finally, we place our results in the context of current spectroscopic surveys to discuss the relevance of the proposed asteroid classes and compositional investigations of individual objects (Section~\ref{sec:implications}).

\section{A brief history of SpeX in the context of asteroid science}
\label{sec:spexscience}

\subsection{The SpeX spectrograph}
\label{sec:spex}

SpeX is a NIR spectrograph and imager used on the 3~m NASA/IRTF on Maunakea (Hawaii) since May 2000. 
The instrument provides a range of observing modes with resolving power up to R$\sim$2000 and wavelength ranges between 0.7 and 5.5\,$\upmu$m. 
An overview of the design, performance and observing techniques of the instrument can be found in \citet{Rayner:2003}, and additional information about the first years of operational experience is provided in \citet{Rayner:2004}. 

Owing to the faint apparent magnitude of most solar system small bodies and the broad absorption features that characterize their spectra, the community of solar system observers mostly uses SpeX in the low-resolution (R$\sim$200) prism mode for spectroscopic characterization of these objects. 
This mode spanning the 0.8--2.5~$\upmu$m wavelength range covers many diagnostic features of ices and minerals enabling classification and compositional investigation of asteroids, comets and Kuiper-belt objects (KBOs). 

Guiding during SpeX observations can be carried out in two different ways: by use of the so-called GuideDog infrared slit viewer on spillover flux from the object in the slit, or in the visible by use of the MORIS CCD guiding camera \citep{Gulbis:2011} combined with a dichroic filter that redirects light below 0.8-$\upmu$m to the camera. 
Since its installation in 2008, MORIS is usually favored for observations of faint objects as it allows guiding on the full signal of the target, instead of the flux surrounding the slit in the NIR. 

\begin{figure}[ht]
   \centering
   \includegraphics[width=0.45\linewidth]{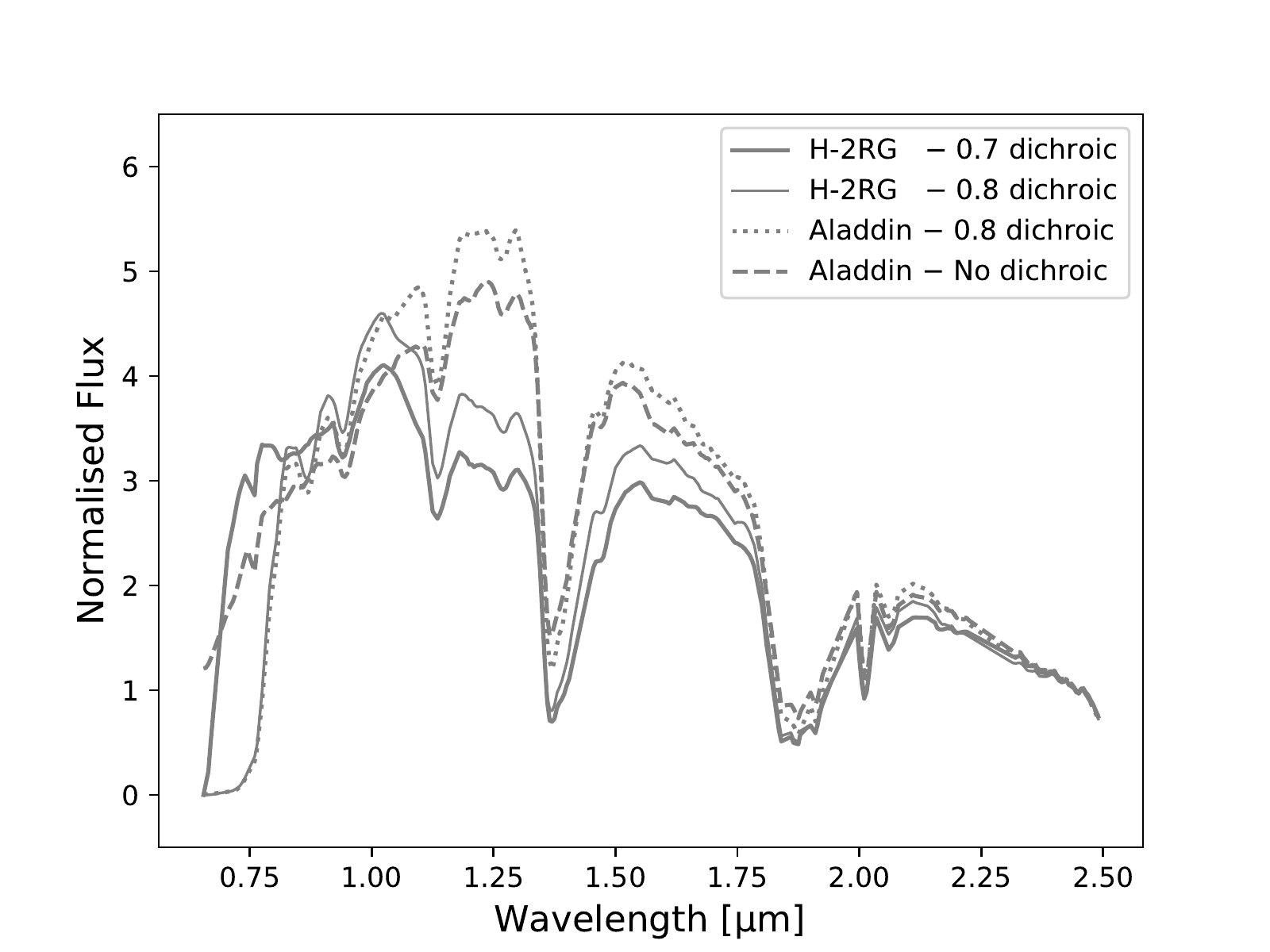}
   \caption{Changes in the throughput of SpeX as consequence of instrumental upgrades. 
   Each curve represents a solar-type spectrum convolved by the optical transmission function and the detector response function for a given detector+dichroic configuration (see Section\,\ref{sec:specvar}). 
   All spectra are normalized to unity at 2.45\,$\upmu$m.
   Even though the changes induced by the upgrades are important, they do not affect asteroid surveys as the process of dividing the asteroid by a stellar spectrum to measure reflectance removes these effects. }
\label{fig:averaged_spectra}
\end{figure}

Two major upgrades occurred during the lifetime of SpeX. 
In 2014 August, the original Raytheon Aladdin 3 1024$\times$1024 InSb array was replaced by a Teledyne 2048$\times$2048 Hawaii-2RG (H-2RG) array. 
In semester 2017A, the 0.8~$\upmu$m cut-on dichroic filter was replaced with a 0.7~$\upmu$m dichroic in order to expand the short end wavelength coverage of the spectrograph when using MORIS. Each of these upgrades led to a significant change in the throughput of the instrument, by modifying the detector response curve and the in-band transmission of light to the spectrograph (Fig.\,\ref{fig:averaged_spectra}). 

\subsection{Large asteroid surveys with SpeX}
\label{sec:smass}

Along its 20 years of operation, more than 1,300 objects in the asteroid belt and near-Earth space have been surveyed by SpeX as part of two large observing programs: the Small Main-Belt Asteroid Spectroscopic Survey (SMASS, \citealt{Xu:1995, Bus:2002, Binzel:2004}) and the MIT-Hawaii Near-Earth Object Spectroscopic Survey (MITHNEOS, \citealt{binzel:2019}). 
These large programs resulted in today's largest NIR spectral database of asteroids, upon which a widely used classification system was built \citep{demeo:2009}. 

Asteroid observations in SMASS and MITHNEOS are alternated with measurements of stars known to be very close spectral analogs to the Sun to remove solar colors from the asteroid spectra by calculating the ratio between the asteroid and the stellar spectra (solar analogs are commonly defined as stars with temperature within $\sim$500\,K and metallicity within a factor 2 of the Sun's, and no close companion; \citealt{Strobel:1996}).
Typically, three stars are chosen to be observed among Hyades 64 and \citet{Landolt:1983}'s stars listed in Table~\ref{tab:sa}. 
These multiple measurements are used to identify possible outlier measurements and to mitigate spectral variability across the measurements by computing a mean stellar spectrum from the three measurements. 
Beyond correcting for solar colors, dividing the asteroid by the star simultaneously cancels most variations due to the varying optical transmission function and detector response function, resulting in a very homogeneous dataset of asteroid reflectance spectra. 
Additional information about the surveys, including details about the different steps of data processing, can be found in \citet{Binzel:2004, binzel:2019}. 


\begin{deluxetable}{llcclcc}
\caption{Solar analog stars used in the SMASS and MITHNEOS spectroscopic surveys.
\label{tab:sa}}
\tablehead{
\colhead{Designation} & \colhead{Identifier} & \colhead{J2000~R.A.} & \colhead{J2000~Dec.} & \colhead{Spt. Type} & \colhead{V$_{\rm mag}$} & \colhead{$N$}
}
\startdata
Hyades 64	     & HD 28099      & 04:26:40.1 & +16:44:49 & G2V${}^{1}$      & 8.1  & 76 \\ 
Land. 93-101	 & HD 11532      & 01:53:18.4 & +00:22:23 & G5V${}^{2}$ & 9.7  & 89 \\ 
Land. 98-978	 & HD 292561     & 06:51:33.7 & -00:11:32 & G3V${}^{2}$     & 10.6 & 71 \\ 
Land. 102-1081   & BD+00 2717    & 10:57:04.0 & -00:13:13 & G5IV${}^{2}$     & 9.9  & 82 \\ 
Land. 105-56	 & BD-00 2719    & 13:38:41.9 & -01:14:17 & G5V${}^{2}$      & 10.0 & 60 \\ 
Land. 107-684    & HD 139287     & 15:37:18.1 & -00:09:50 & G3V${}^{2}$    & 8.4  & 31 \\ 
Land. 107-998    & BD+00 3383    & 15:38:16.3 & +00:15:22 & G3IV${}^{2}$     & 10.5 & 23 \\ 
Land. 110-361    & TYC 447-508-1 & 18:42:45.0 & +00:08:05 & G2${}^{3}$      & 12.4 & 64 \\ 
Land. 112-1333   & BD-00 4074    & 20:43:12.0 & +00:26:13 & G2V${}^{2}$      & 10.0 & 29 \\ 
Land. 113-276    & BD-00 4251B   & 21:42:27.4 & +00:26:20 & G5V${}^{2}$      & 9.1  & 40 \\ 
Land. 115-271    & BD-00 4557    & 23:42:41.8 & +00:45:13 & G2V${}^{2}$      & 9.7	& 63 \\ 
\enddata
 \tablecomments{`Land.' refers to stars selected from \citet{Landolt:1983}. $N$ is the number of available spectral measurement in our dataset for each solar analog. Spectral types from: ${}^{1}$\citet{Keenan:1989}, ${}^{2}$\citet{Drilling:1979}, ${}^{3}$estimated based on optical and near-infrared colors from \citet{Landolt:1992} and \citet{Cutri:2003}.}
\vspace{2mm}
\end{deluxetable}

\section{Investigation of uncertainties and systematics in asteroid surveys} \label{sec:errors}

\subsection{Intrinsic uncertainties}
\label{sec:specvar}

The consistent set of 11 solar analogs used over 20 years of SMASS and MITHNEOS observations provides a unique dataset of spectral measurements to explore the evolution of the throughput of SpeX over time. 
Here, we use those stellar measurements to investigate spectral variability in SpeX observations, derive reliable uncertainties on the measured spectral slopes, and search for possible correlations between the measurements, instrumental effects and weather conditions.  
Whereas the very first SpeX observations in SMASS date back to September 2000, shortly after the commissioning of the instrument (April 2000), only stellar spectra collected in November-December 2002 and after December 2004 are still available today on the internal SMASS server (raw data for all observations, including pre-2004, can be retrieved from the IRTF Legacy Archive\footnote{\url{http://irtfdata.ifa.hawaii.edu/}}). 
Consequently, the data analysed in this work consist of the complete set of stellar spectra acquired in November-December 2002 and between December 2004 and November 2019 
(628 spectra in total). 

\begin{figure}[ht]
   \centering
   \includegraphics[width=0.45\linewidth]{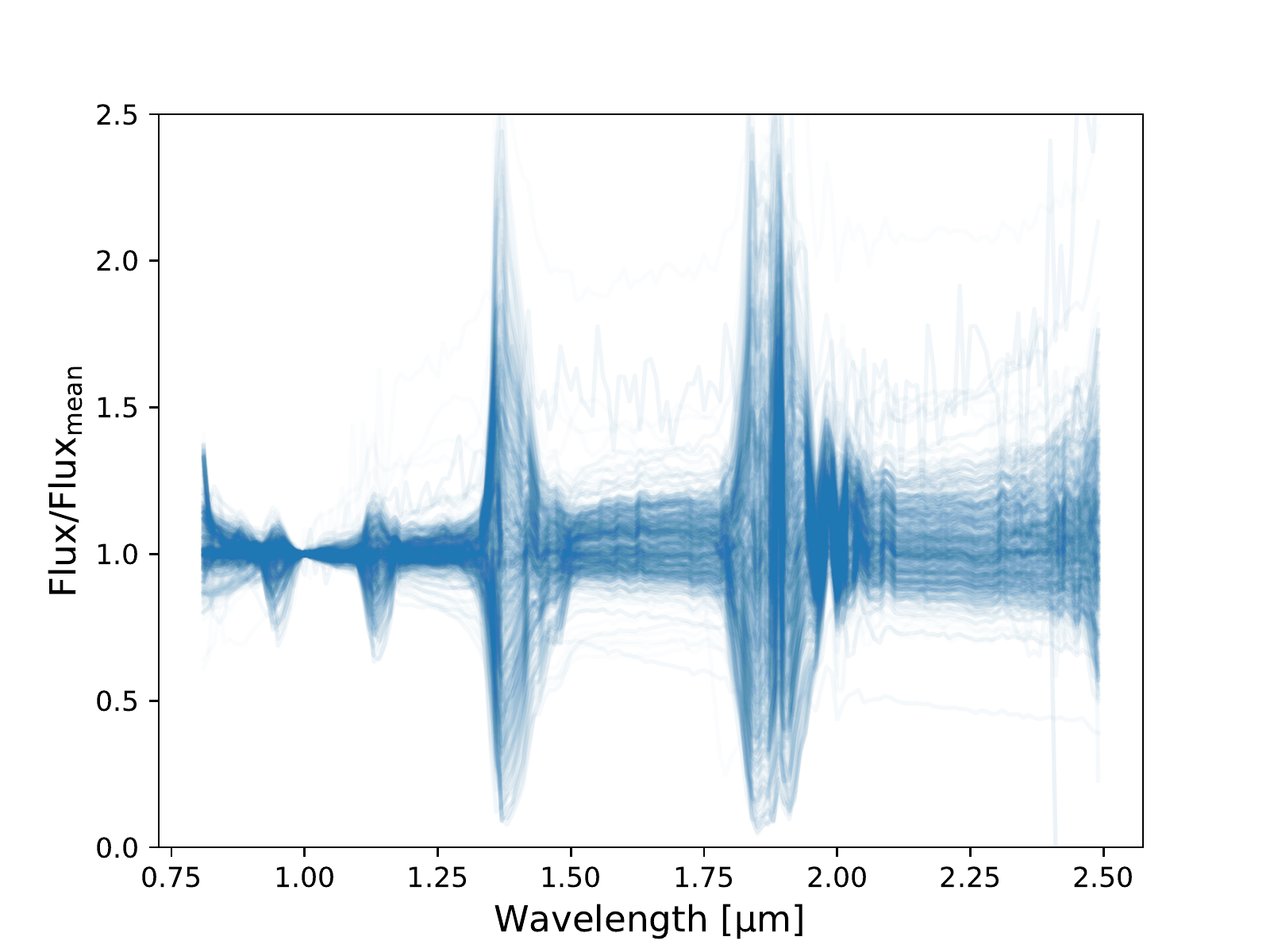}
   \includegraphics[width=0.45\linewidth]{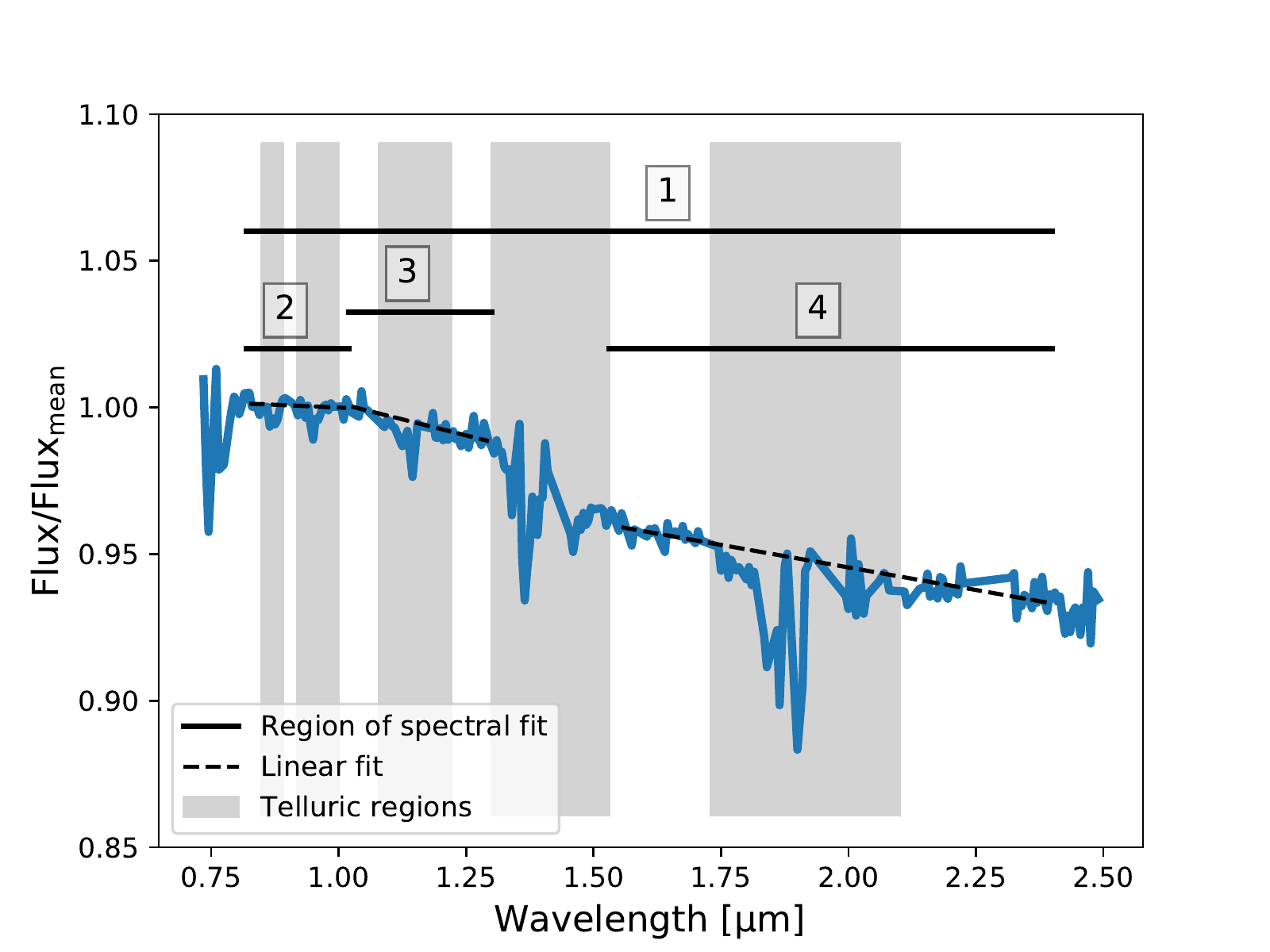}
   \caption{Illustration of the method used to investigate spectral variability in the SMASS and MITHNEOS surveys. 
   {\bf Left:} Complete set of 628 solar-type spectra  analysed in this work. Each spectrum was divided by a mean stellar spectrum computed from the full set of measurements taken under the same detector+dichroic configuration. 
   {\bf Right:} For each individual spectrum (here Land~107-998 acquired on 05-17-2013), 
   linear fits (dotted line) were performed over the 0.80--2.40~$\upmu$m, 0.80--1.02~$\upmu$m, 1.02--1.30~$\upmu$m and 1.55--2.40\,$\upmu$m wavelength intervals (labelled 1, 2, 3 and 4 on the figure), avoiding regions of strong atmospheric bands. } 
\label{fig:splashy}
\end{figure}

As a first step, temporal variability in our dataset was investigated by dividing each of the 628 spectra by an average stellar spectrum, and then by measuring spectral slope after normalizing the spectrum to unity at 1\,$\upmu$m. 
Slope measurements were performed over four wavelength ranges: the overall
0.80--2.40\,$\upmu$m interval, as well as  
0.80--1.02\,$\upmu$m, 
1.02--1.30\,$\upmu$m and 
1.55--2.40\,$\upmu$m, 
excluding regions of strong telluric absorptions within these intervals. 
These ranges were defined based on the position of the inflections of the spectra near 1.0 and 1.3\,$\upmu$m (Fig.\,\ref{fig:splashy}).  

\begin{figure}[ht]
   \centering
   \includegraphics[width=0.8\linewidth]{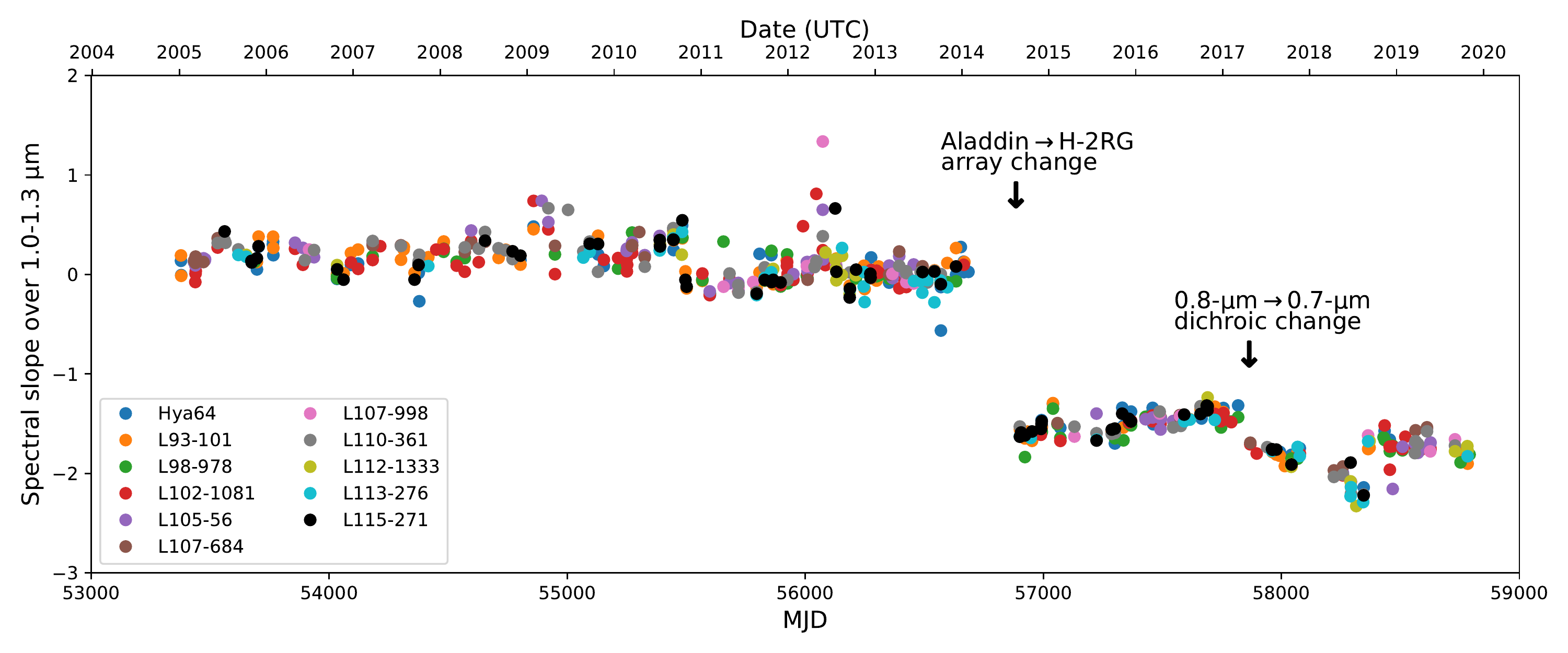}
   \includegraphics[width=0.8\linewidth]{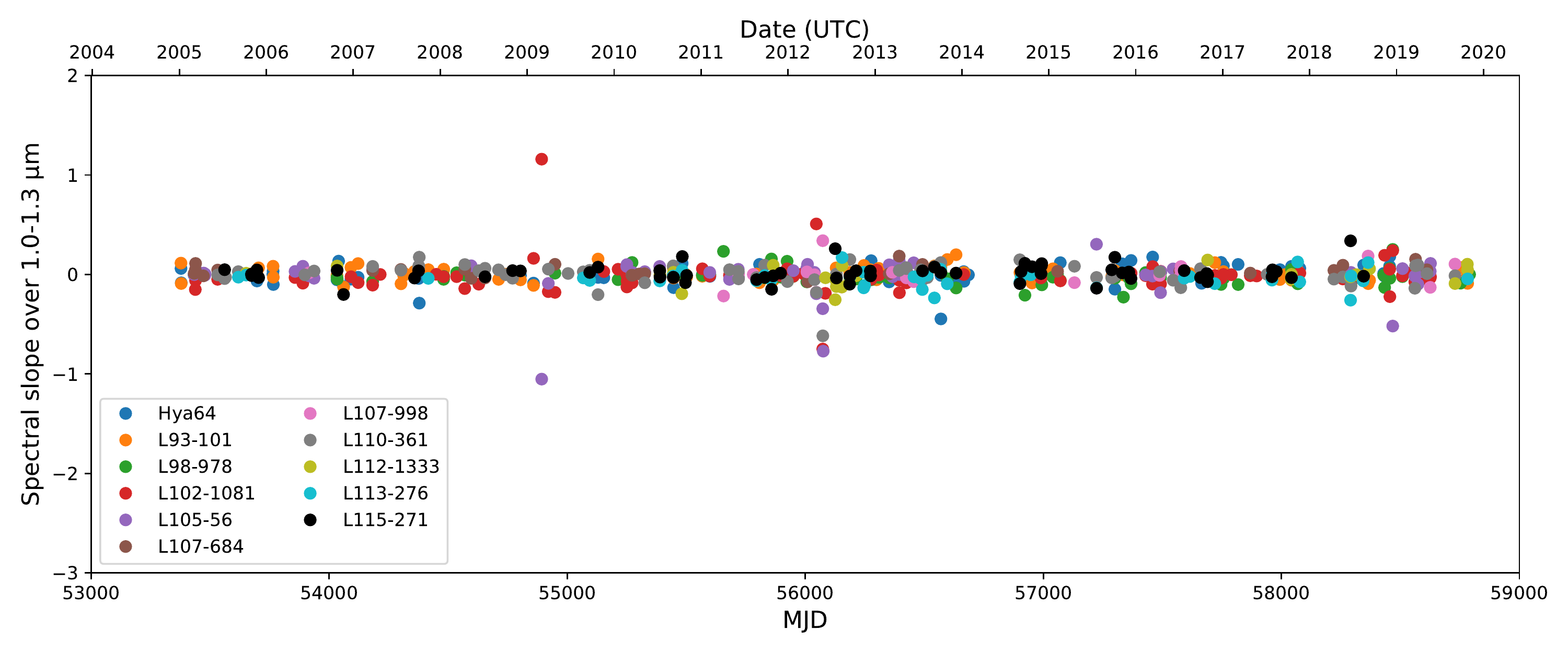}
\caption{{\bf Top:} Long-term variability is created by instrument updates and telescope maintenance among spectral slopes of solar analogs collected from 2004 to 2019 by the SMASS and MITHNEOS surveys. Here, slope measurements calculated from 1.0 to 1.3\,$\upmu$m are shown, because this wavelength range exhibits the largest variability in our dataset (data acquired before 2004 are not plotted for better readability). All spectra were divided by a mean stellar spectrum built from all data collected before 2014 April (before the Aladdin to H-2GR array change). Notable changes are due to an array change in 2014, and a dichroic change in 2017. Additional variability is observed with periodicity of months or a year, and is due to telescope maintenance and cleaning (see Section\,\ref{sec:specvar}). These variations do not affect the slope of the final asteroid spectra because they are corrected in the
asteroid-star division. 
{\bf Bottom:} A 5-day average is applied to correct for long-term variability. From this
corrected data, we can calculate the standard deviation to determine the overall slope variability of the SMASS and MITHNEOS surveys (Section\,\ref{sec:specvar}), and investigate correlations between measurements, observing conditions and instrumental configuration (Section\,\ref{sec:correlations}).}
\label{fig:temporal_changes}
\end{figure}

The upper panel of Fig.\,\ref{fig:temporal_changes} shows the temporal variation of spectral measurements from 1.02 to 1.30\,$\upmu$m. 
This wavelength range is the most sensitive to instrumental upgrades and illustrates well spectral variability in our dataset. 
Notable spectral variations are due to array change in 2014 and dichroic change in 2017. 
Additional periodic variability is observed over timescales of months or a year, and mostly due to telescope maintenance and cleaning: 
recoating the primary mirror in 2012, monthly CO$_2$ cleaning of the primary and a wet wash about once a year. 
Although these effects largely influence the throughput of the telescope over time, they are cancelled in asteroid surveys by the process of dividing the asteroid spectra by the stellar ones measured on the same night through the same instrumental configuration. 
To derive reliable uncertainties for spectral slope measurements of asteroids, we therefore needed to remove these long-term effects in a similar way as performed in asteroid surveys. 
This was achieved by calculating a running average spectral slope value over a 5-day window, and by subtracting this average from the corresponding slope measurement. 
The resulting corrected measurements are shown in the bottom panel of Fig.\,\ref{fig:temporal_changes}: the flat distribution indicates that most long-term effects have been removed, and that the distribution is now dominated by the statistical uncertainty. 
This is further verified in Section\,\ref{sec:correlations}, where possible remaining systematics are investigated. 

\begin{figure}[ht]
   \centering
   \includegraphics[width=0.45\linewidth]{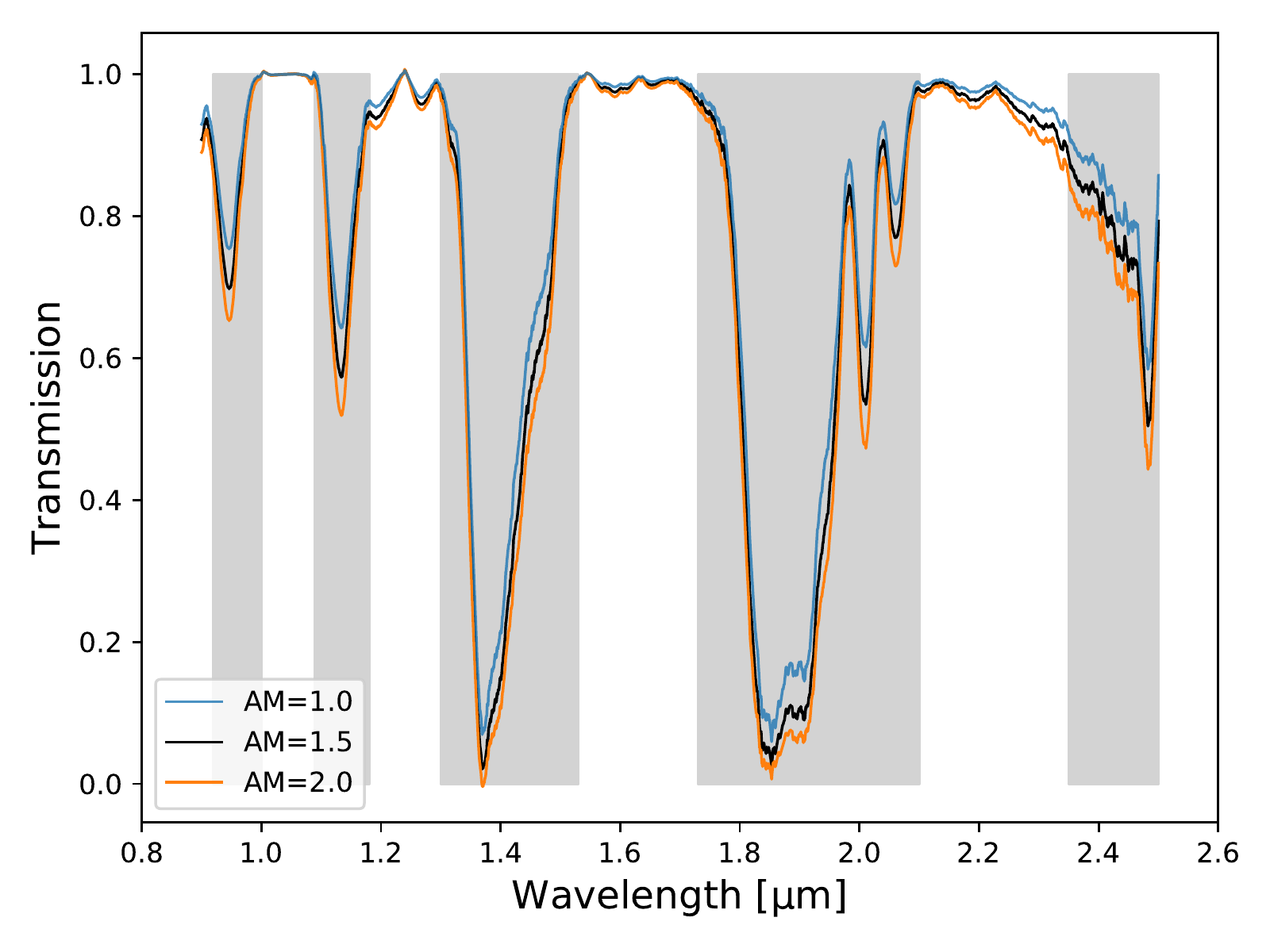}
   \caption{Infrared spectra of the atmospheric transmission above Maunakea generated by use of the ATRAN modelling software \citep{Lord:1992} for 1.6\,mm precipitable water vapour at 1.0 (blue), 1.5 (black) and 2.0 (orange) air mass (AM). The grey regions indicate spectral intervals of strong telluric absorption. Division of the 1.5 AM transmission spectrum by the 1.0 one results in a negative spectral slope of -1.4\%\,$\upmu$m${}^{-1}$ over the 0.80--2.40\,$\upmu$m wavelength range. Dividing the 2.0 spectrum by the 1.0 one results in a negative slope of -2.6\%\,$\upmu$m${}^{-1}$. We therefore estimate that the ATRAN atmospheric correction performed in SMASS and MITHNEOS accounts for a slope correction of $\sim$-0.3\%\,$\upmu$m${}^{-1}$ per 0.1 unit AM. A similar correction was applied to our stellar dataset.} 
\label{fig:atran}
\end{figure}

Atmospheric correction of the asteroid and stellar spectra in SMASS and MITHNEOS is performed by use of the ATRAN modelling software \citep{Lord:1992} in order to remove telluric absorption features from the spectra. 
Stellar spectra internally available today on the SMASS database however were recorded before this step of the reduction pipeline, meaning that atmospheric effects are still present in the spectra. 
In order to get a reliable estimate of the ATRAN correction performed in the final steps of the reduction process, we generated an atmospheric transmission model at Maunakea for air masses of 1.0, 1.5 and 2.0, and precipitable water vapor of 1.6\,mm (Fig.\,\ref{fig:atran}). 
Dividing the two high air mass transmission spectra by the low air mass one results in a negative slope of -0.3\%\,$\upmu$m${}^{-1}$ per 0.1 unit air mass. 
This value was used to correct for air mass differences between the various spectra in our dataset. 
Because precipitable water vapor was not found to contribute to systematic slope change in our dataset (Section\,\ref{sec:correlations}), we did not perform any further correction for this parameter. 
Additional information about near-infrared extinction curves at Maunakea can be found in \citet{Tokunaga:2002}.

\begin{figure}[ht]
   \centering
   \includegraphics[width=0.65\linewidth]{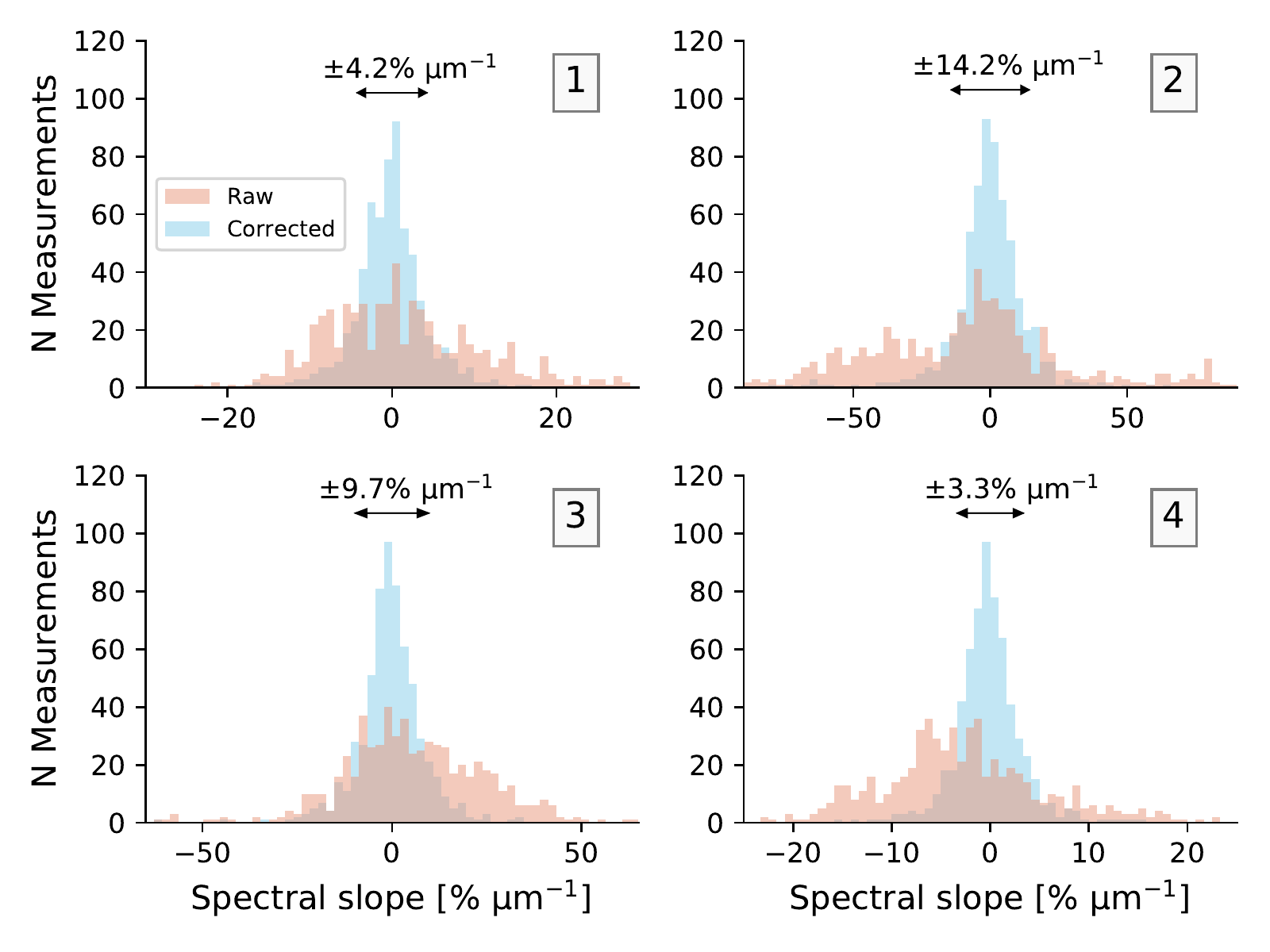}
   \caption{
   Histograms of slope measurements from the complete set of stellar spectra. 
   The number at the upper-right corner of each panel indicates the wavelength interval over which the slope was measured (as defined in Fig.\,\ref{fig:splashy}). 
   The red histograms correspond to stellar spectra divided by an average computed by use of all measurements collected under the same array+dichroic configuration. 
   The blue ones were corrected for long-term variability as described in Section~\ref{sec:specvar}. 
   The slope uncertainty is calculated as the 1$\upsigma$ deviation of the corrected measurements and indicated in each panel.}
\label{fig:L107998}
\end{figure}

Finally, uncertainties on spectral slope measurements were computed as the standard 1$\upsigma$ deviation with respect to the mean. 
Five out of the 628 measurements found to be 4$\upsigma$ inconsistent with the overall dataset were rejected from the calculation. 
The derived values for the four wavelength ranges are provided in Fig.\,\ref{fig:L107998}. 
On the overall 0.80--2.40~$\upmu$m wavelength range, where most spectral slope measurements of asteroids are reported in the literature, the slope uncertainty is 4.2\% per micron. 
Implications of the derived uncertainty for current classification systems and compositional investigations of asteroids are discussed in Section~\ref{sec:implications}. 

\subsection{Systematic errors}
\label{sec:correlations}

As a next step, we searched for possible correlations between the corrected slope measurements and observing circumstances (both weather conditions and instrumental configurations) to identify possible sources of systematics in asteroid surveys. 
Information on air mass, focus position, seeing, precipitable water vapor, humidity, wind speed and direction, and air temperature, were collected directly from the headers of the original FITS spectral images whenever recorded.\footnote{Similarly, all observing circumstances for asteroid spectra in the SMASS and MITHNEOS database, including observing geometry (phase angle, heliocentric and geocentric distances), are now publicly available at \url{http://smass.mit.edu/minus.html}.} 
We also recalculated the parallactic angle for each observation in order to investigate possible effects of differential atmospheric refraction on the measurements \citep{Filippenko:1982} -- the slit being always aligned with the equatorial North-South direction in SMASS and MITHNEOS 
for survey configuration consistency and time efficiency. 

\begin{figure}[ht]
   \centering
   \includegraphics[width=0.45\linewidth]{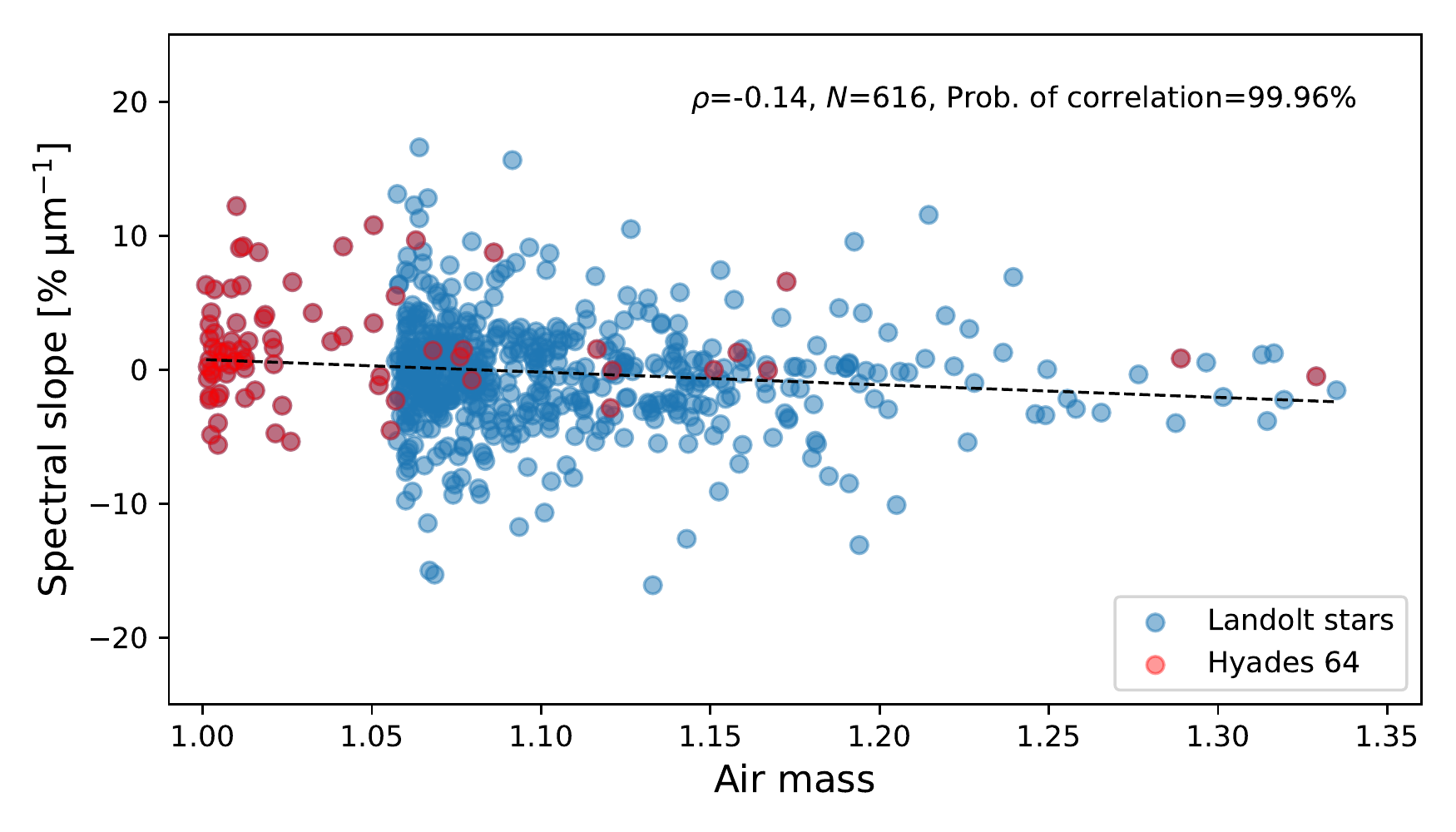}
   \includegraphics[width=0.45\linewidth]{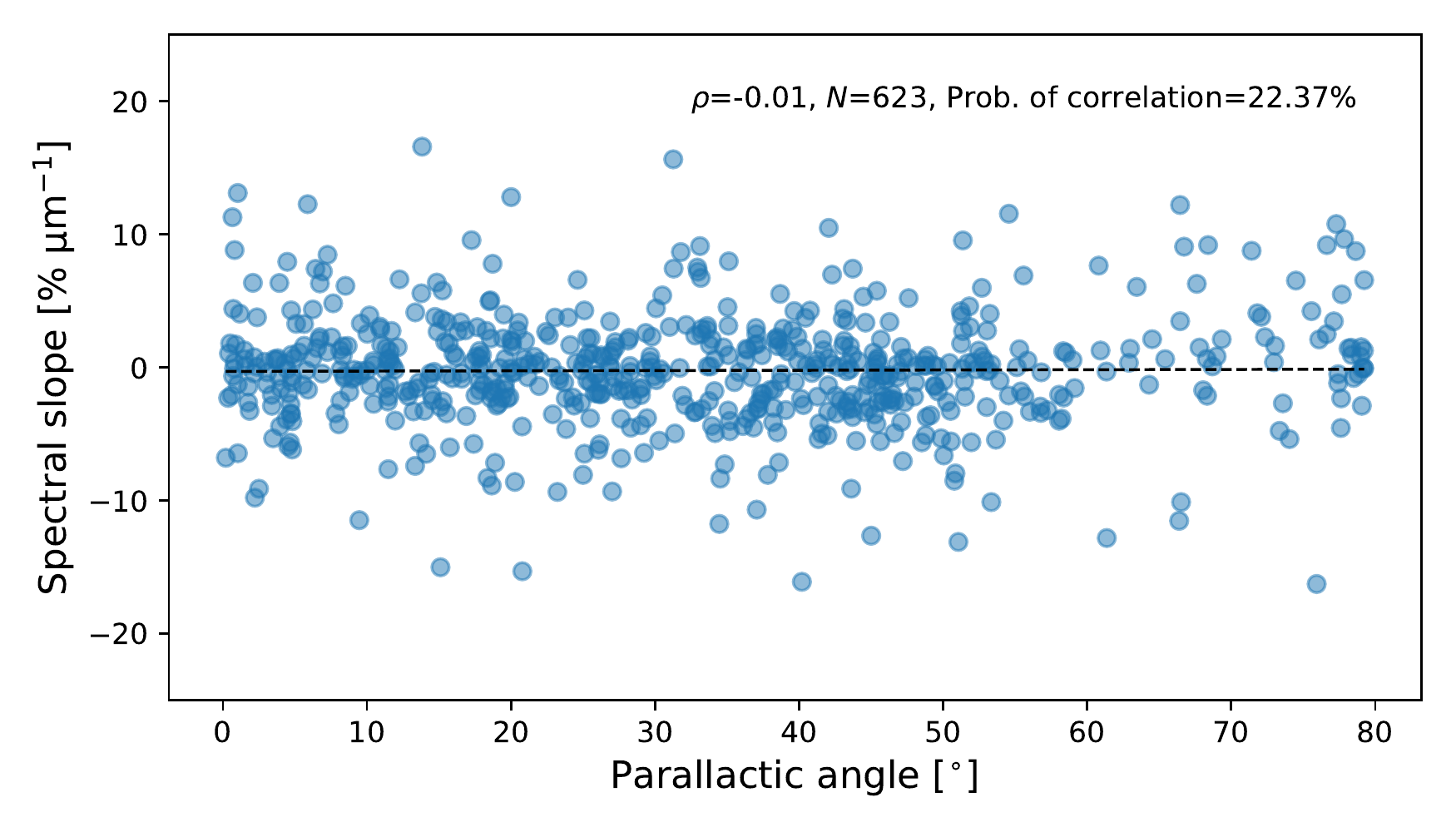}
   \caption{Search for correlations between 0.8--2.4\,$\upmu$m spectral slope measurements and observing conditions. 
   The dotted line is a linear fit to the data. 
   The Spearman rank-order correlation coefficient ($\rho$), the number of observations ($N$), and the corresponding probability of correlation (Section\,\ref{sec:correlations}) are indicated in each panel. 
   Five out of the 628 measurements found to be 4$\upsigma$ inconsistent with the overall dataset were rejected from the calculation of the correlation.  
   Only air mass affects the measurements, with -0.92\%\,$\upmu$m${}^{-1}$ slope variation per 0.1 unit air mass (left panel; only measurements within 1.0--1.35 air mass are shown, see Fig.~\ref{fig:app_A1} of Appendix~\ref{sec:app_A} for the complete range of air masses). Other parameters (e.g. the absolute value of the parallactic angle, right panel) are uncorrelated to the measurements (see further discussion in Section\,\ref{sec:correlations}). 
   The peculiar distribution of air mass values arises from the apparent position of the stars as seen from Hawaiian latitudes (+20 degree North). 
   Landolt stars (in blue) never reach air masses below $\sim$1.05 due to their location on the celestial equator, unlike Hyades 64 (red) that is located near +17 degrees declination and therefore transiting near the zenith. 
   The position of the stars also explains why the parallactic angle is never larger than $\sim$80 degrees.
   Similar plots for each recorded observing condition and instrumental parameters are provided in Fig.~\ref{fig:app_A1} of Appendix~\ref{sec:app_A}.} 
\label{fig:parang}
\end{figure}

\begin{figure}[ht]
   \centering
   \includegraphics[width=0.45\linewidth]{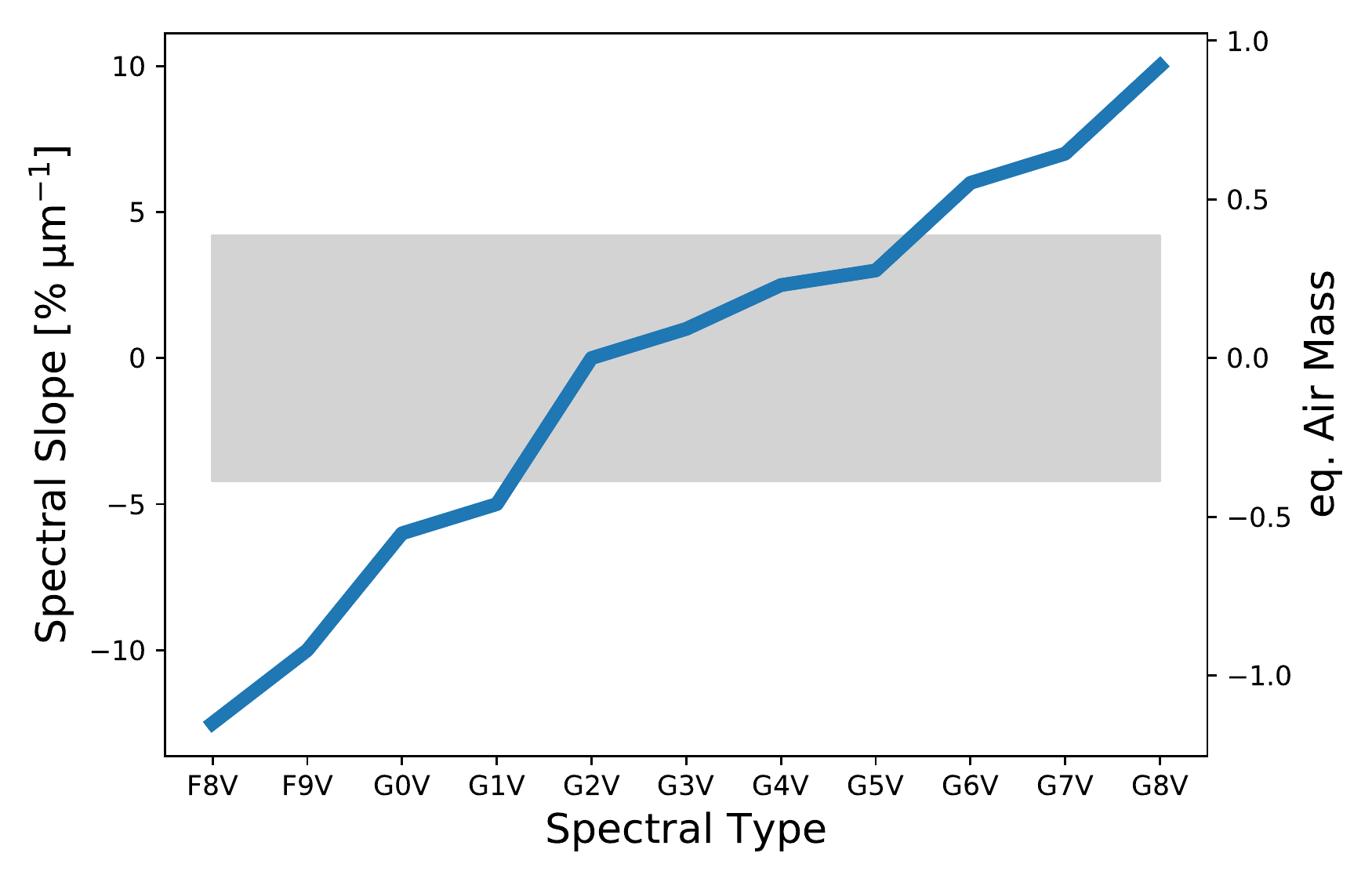}
   \caption{NIR spectral slope of late FV and GV-type stellar spectra divided by a solar spectrum, converted from their (R-K) colors from \citet{Pecaut:2013} using the Synphot/STSDAS software (\url{www.stsci.edu/institute/software_hardware/stsdas}; 2MASS magnitudes in \citet{Pecaut:2013} were converted to Bessel using equations from \citet{Carpenter:2001} for compatibility with Synphot). The grey region indicates the $\pm$4.2\%\,$\upmu$m${}^{-1}$ statistical uncertainty with respect to 0 (solar spectrum). A change in stellar type within a reasonable range (e.g., G2V to G5V) produces a slope change of a few percents per micron. This is equivalent to the effect of a few tenths of unit air mass difference between the science target and the solar analog used to divide the spectrum. Even within a given spectral type, stars can exhibit some spread in spectral slope. Stars used for color correction of the asteroid spectrum must therefore be carefully chosen to obtain a consistent set of asteroid reflectance spectra. In SMASS and MITHNEOS, solar analogs vary within less than 2\%\,$\upmu$m$^{-1}$ compared to the sun (Fig.\,\ref{fig:stellarVariability}).}
\label{fig:am_vs_spt}
\end{figure}

\begin{figure}[ht]
   \centering
   \includegraphics[width=0.45\linewidth]{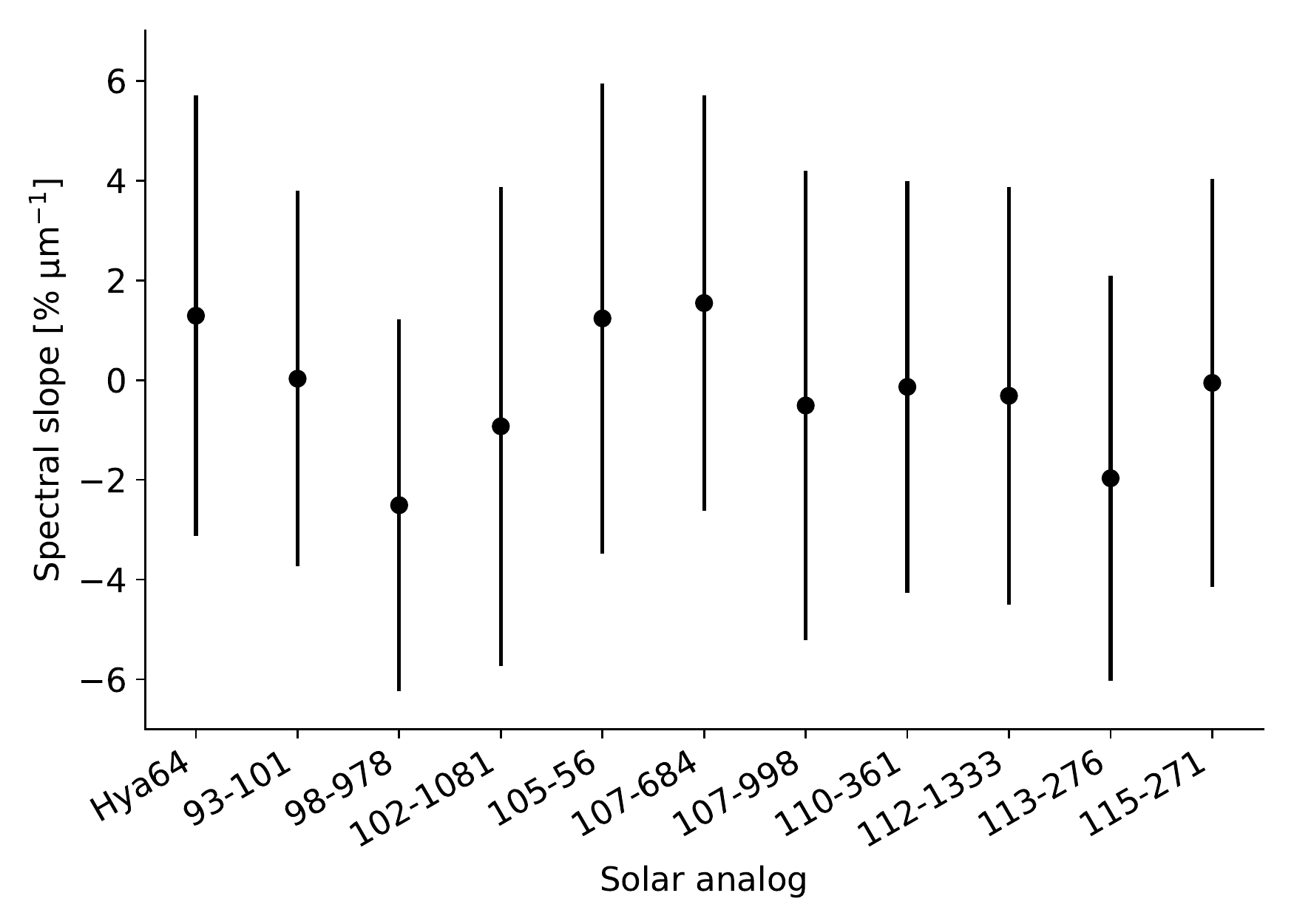}
   \caption{Star-to-star spectral slope variability in our dataset. The 1$\upsigma$ deviation of stellar averages is 1.4\%\,$\upmu$m${}^{-1}$, well within the overall 4.2\%\,$\upmu$m${}^{-1}$ uncertainty in our dataset. The choice of the solar analog from Table\,\ref{tab:sa} used to calibrate asteroid spectra therefore does not influence significantly the measured reflectance slope of the asteroid.} 
\label{fig:stellarVariability}
\end{figure}

Statistical correlations between spectral slope measurements and observing parameters were searched by means of the Spearman rank test. This test returns a parameter $\rho$, comprised between -1 and +1, calculated as the ratio between the covariance of two rank variables (numerator) and the product of their standard deviations (denominator). This parameter describes how well the relationship between the variables can be described by a {\it monotonic} function (-1 and +1 imply an exact monotonic relationship whereas 0 implies no correlation). The Spearman test slightly differs from the Pearson test, which assesses how well two variables can be described by a {\it linear} relationship. 
Here, we preferred Spearman over Pearson as some of the analysed parameters are not expected to vary at a constant rate. 
From $\rho$ and the number of measurements $N$, we then computed a probability of correlation ($P)$ for each parameter as $1-p$, where $p$ is the two-sided p-value of the test giving the probability that the observed distribution was drawn from a random distribution. 
By doing so, spectral slope was found to either correlate or anti-correlate with air mass over each of four wavelength ranges. 
Specifically, spectral slope correlates with air mass over 0.80--1.02\,$\upmu$m with $P=99.43\%$ and then anti-correlates over the 1.02--1.30\,$\upmu$m and 1.55--2.40\,$\upmu$m ranges with $P=92.04\%$ and $P>99.99\%$, respectively. 
On the overall 0.80 to 2.40\,$\upmu$m interval, the same anti-correlation is observed with $P=99.96\%$ when considering measurements within 1.0 to 1.3 air mass, and $P>99.99\%$ when including measurements beyond that range (Fig.\,\ref{fig:parang}). 
The induced spectral change on the full wavelength interval is -0.92\%\,$\upmu$m$^{-1}$ per 0.1 unit air mass. 
Most asteroid surveys, including SMASS and MITHNEOS, are designed to operate mostly in the 1.0 to 1.3 air mass range (97.5\% of the measurements analysed here were acquired within that range of values), implying that air mass variation is responsible for no more than a 2.8\%\,$\upmu$m$^{-1}$ variability of spectral slope measurements (i.e. well within the overall 4.2\%\,$\upmu$m$^{-1}$ variability in our dataset). 
This effect is of the same order as slope change from average G2V to G5V stars (Fig.\,\ref{fig:am_vs_spt}). It is known, however, that even within a given spectral type, stars can exhibit some spread in spectral slope. 
Solar analogs should therefore be carefully chosen to be as spectrally similar as possible to the Sun over the range of observed wavelengths.
In SMASS and MITHNEOS, stars vary in slope by only $\pm$2\%\,$\upmu$m$^{-1}$ compared to the sun (Fig.\,\ref{fig:stellarVariability}). 

Aside from air mass, none of the other instrumental and weather parameters explored in this work, including parallactic angle, seeing and humidity, were found to induce systematic errors in slope measurements at the $>$2$\upsigma$ confidence level (Fig.\,\ref{fig:parang} and Appendix~\ref{sec:app_A}). 
In their analysis of the performances of SpeX during its first years of operation, \citet{Rayner:2004} noted that observing away from the parallactic angle can induce a slope variation of about 7\% across the 2.5\,$\upmu$m baseline. However, the authors used a narrower slit of 0.3", against 0.8" in our case (i.e., larger than the average seeing of $\sim$0.5" at Maunakea; e.g. \citet{Racine:1989} and Appendix~\ref{sec:app_A}). We therefore stress that our conclusions are only valid in the case of observations conducted with the wider 0.8" slit, the most commonly used slit in spectroscopic surveys of asteroids. 
The 0.3" slit is rarely used for asteroid spectroscopy because these objects do not have narrow spectral features, so there is no need for higher spectral resolution, 
especially at the cost of losing light for a faint object. 
It is further likely that observations in our dataset acquired away from the parallactic angle in poor seeing conditions ($>$0.8") might be affected by differential refraction, but the number of measurements acquired in such conditions and for which seeing values are recorded is insufficient to detect any trend.

We then searched for star-to-star variability and temporal variations of individual stars in our dataset. Fig.\,\ref{fig:stellarVariability} shows the average spectral slope value over 0.80--2.40\,$\upmu$m for the 11 solar analogs: 
all are found to be spectrally similar within errorbars, with 1.4\%\,$\upmu$m$^{-1}$ deviation between stellar averages and a 4.4\%\,$\upmu$m$^{-1}$ difference between the two most extreme averages. 
Scattering of spectral slope measurements does not vary significantly from star to star, suggesting that none is spectrally variable to the level of precision of our measurements. 

\begin{figure}[ht]
   \centering
   \includegraphics[width=0.45\linewidth,align=c]{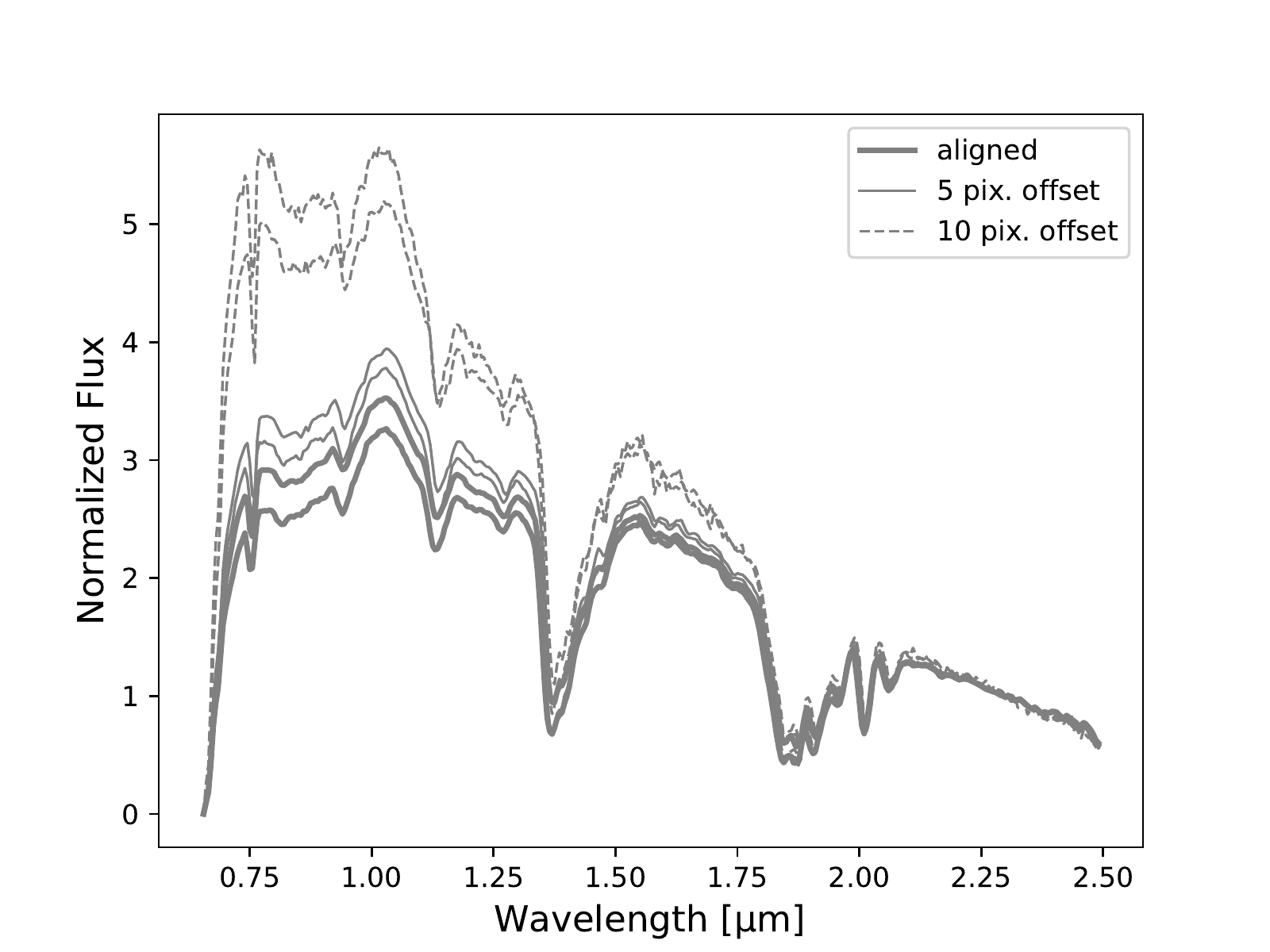}
   \includegraphics[width=0.135\linewidth,align=c]{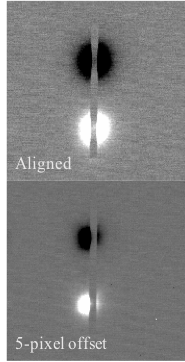}
   \caption{Slit misalignment can be a major source of slope uncertainty in small body surveys. Here, spectra of Land~105-56 ({\it left}) were acquired on two different nights with the star deliberately being alternatively aligned and off-centered with respect to the slit (as shown by the guider images; {\it right}). 
   A 5 pixel offset induces a spectral bluing of 10.2\% on the divided and normalised spectra.} 
\label{fig:align}
\end{figure}

Slit alignment was found to constitute a possible major source of error in spectral slope measurements. 
This was investigated by acquiring spectroscopic observations of stars deliberately aligned and then misaligned on the slit (Fig.\,\ref{fig:align}). 
The off-centered spectrum was then divided by the aligned one, and normalized to unity at 1\,$\upmu$m. 
By doing so, a 5-pixel offset was found to induce a spectral bluing of 10.2\% on the stellar spectrum. 
The overall 4.2\%\,$\upmu$m$^{-1}$ uncertainty in our dataset thus demonstrates that the typical centering accuracy in the SMASS and MITHNEOS surveys must be better than 5 pixels. The best prevention for slit misalignment is to set the target up properly at the start of the observation. 
In SMASS and MITHNEOS, the MORIS guide box is centered at the beginning of the night on the position of the SpeX slit. 
At the start of each target observation, an image in GuideDog is taken to view light surrounding the slit. 
If the target is not properly aligned, the guide box is adjusted at the pixel level and the slit is reimaged to ensure the target is aligned.
Effects due to slit misalignment of the solar analogs are further mitigated by the acquisition of multiple calibration stars allowing the identification and rejection of outlier stellar measurements. 
Additional sanity checks could consist of re-aligning the target on the slit several times during the observations. In the MITHNEOS survey we generally break each asteroid target observation into sets of 20 minutes. When guiding with MORIS was first available, the slit would be re-imaged in GuideDog after each set to ensure object alignment. Because MORIS guiding and IRTF non-sidereal tracking in general is excellent, adjustments at mid-observation (generally throughout a one-hour period) were not needed, and the standard survey procedure is now to perform alignment checks during target set up only. 

Finally, we investigated whether the use of MORIS improves (or diminishes) the accuracy of auto-guiding and object alignment on the slit during the observations. 
In our dataset, the 1$\upsigma$ deviation of slope measurements from spectra obtained while guiding with MORIS is 4.3\%\,$\upmu$m$^{-1}$, whereas for data acquired while guiding with GuideDog it is 4.0\%\,$\upmu$m$^{-1}$. 
The statistical significance of this finding was evaluated by means of Levene's test, which allows testing the null hypothesis that the variances of the two groups of measurements are equal. 
The derived high p-value of the test ($p$=0.20) clearly indicates that the measured difference is statistically insignificant, meaning that we cannot draw any conclusion on to whether the use of MORIS actually influences the accuracy of slit alignment.

\section{Implications for asteroid surveys} \label{sec:implications}

While taxonomic systems of asteroids (e.g., \citealt{Tholen:1984, Bus:2002, demeo:2009}) mostly rely on absorption features to distinguish between spectral classes of objects, several classes can only be differentiated through the slope and overall shape of the continuum of their reflectance spectra. 
This is the case, for instance, for distinguishing among weakly-featured B-, C-, X- and D-type asteroids, as well as between S and Q types, although Q-type asteroids usually exhibit deeper silicate absorption bands than the S types. 
Likewise, most outer solar system objects, including small KBOs, comets and Centaurs, are spectrally featureless in the visible and NIR and can only be distinguished in these wavelengths by means of their spectral slope. 

The uncertainties derived in this work provide a direct test for the use of spectral slope measurements as a way to differentiate between compositional classes of small bodies. 
For instance, in the asteroid taxonomic system of \citet{demeo:2009}, the weakly-featured B-, C-, X- and D-types, with mean spectral slopes of -8.9, 14.9, 22.1 and 39.9\%\,$\upmu$m$^{-1}$ over the 0.8--2.45\,$\upmu$m wavelength range, respectively (-9.9, 12.3, 27.3 and 61.2\%\,$\upmu$m$^{-1}$ over the overall 0.45 to 2.45\%\,$\upmu$m$^{-1}$), 
can be reliably differentiated from one another considering the 4.2\%\,$\upmu$m$^{-1}$ accuracy on slope measurements. 
Q and S-type classes, on the other hand, have mean spectral slopes of 21.1 and 14.2\%\,$\upmu$m$^{-1}$, respectively (13.6 and 20.7\%\,$\upmu$m$^{-1}$ over 0.45 to 2.45\%\,$\upmu$m$^{-1}$): 
spectral slope differentiation between these two classes can therefore be ambiguous, and a combined slope and band analysis, or a Principal Component Analysis (PCA), would ideally be required for reliable classification of objects in these two classes. 
Average spectral slope and associated 1$\upsigma$ spread of taxonomic classes in the \citet{demeo:2009}'s classification scheme are provided in Table\,\ref{tab:taxons}. 
For the most part, taxonomic assignment is performed through PCA on slope-removed spectra in this classification scheme, meaning that, in many cases, slope uncertainty does not influence the classification. 
Some steps of the classification flowcharts however do use the calculated slope value to differentiate between some classes of objects (see Appendix~B and C of \citealt{demeo:2009}): 
uncertainties derived in this work will help evaluating the reliability of these classification steps in future works on taxonomic assignment.

\begin{figure}
   \centering
   \includegraphics[width=0.45\linewidth]{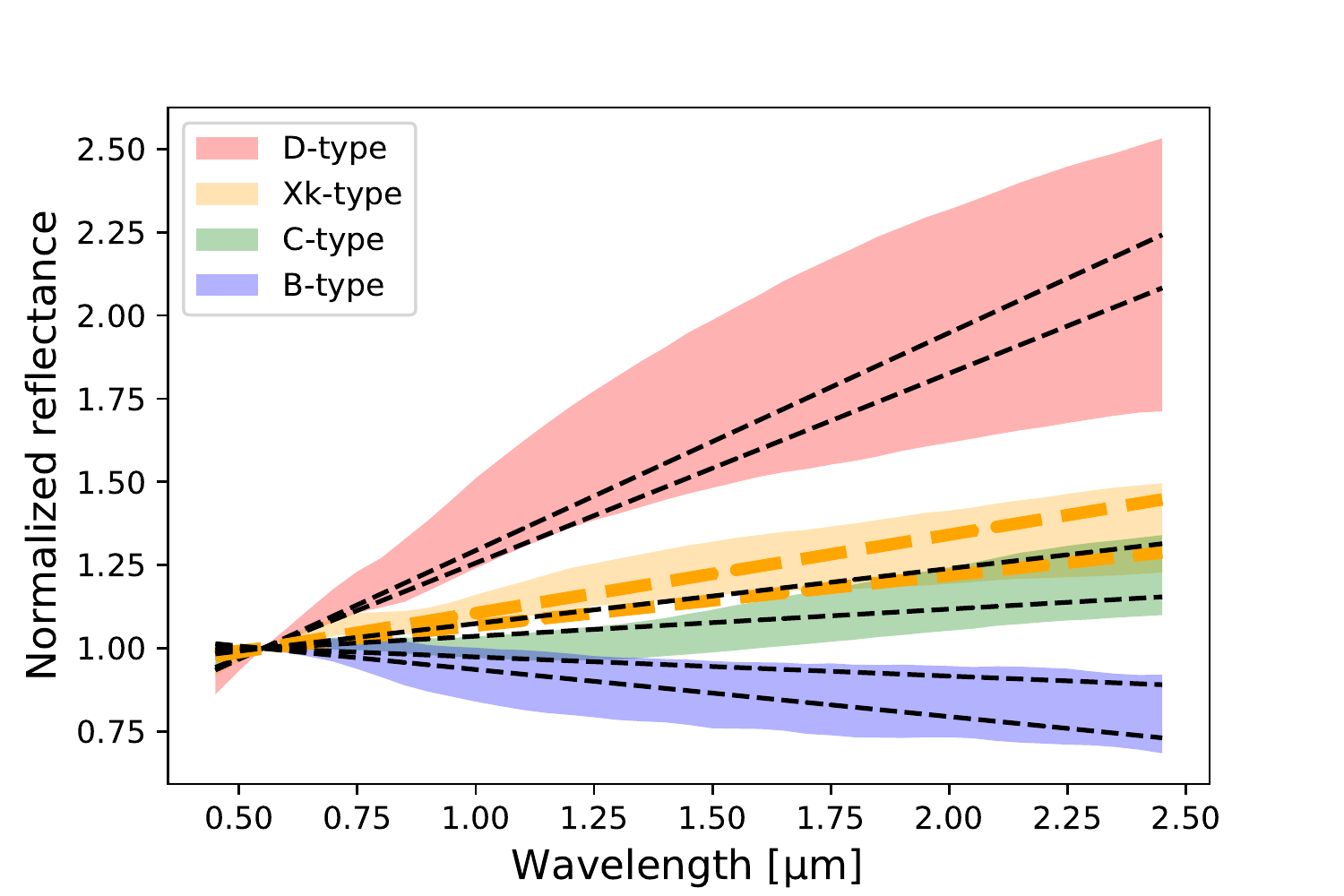}
   \includegraphics[width=0.45\linewidth]{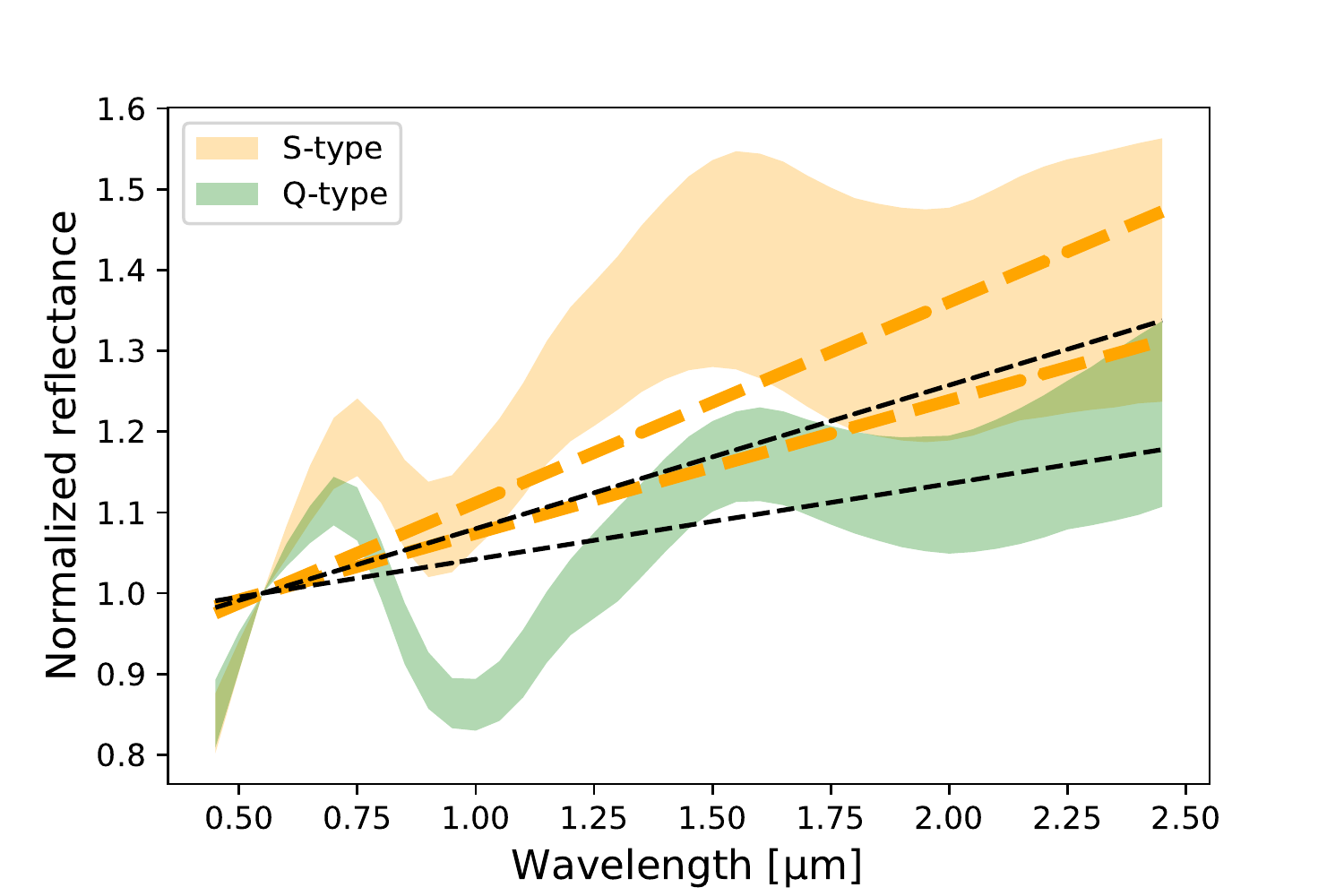}
   \caption{1$\upsigma$ spectral range of some of the major taxonomic classes of asteroids from \citet{demeo:2009}. 
   All spectra are normalised to unity at 0.55\,$\upmu$m.
   The dotted lines indicate linear fits to the spectra plus-minus the statistical 4.2\%\,$\upmu$m$^{-1}$ slope uncertainty derived in this work (a different color is used for the Xk-type and S-type classes for clarity). 
   Linear spectral fits were calculated in the same way as in \citet{demeo:2009}'s taxonomic classification tool (\url{http://smass.mit.edu/busdemeoclass.html}): all data points from 0.45 to 2.45\,$\upmu$m were fitted using the same statistical weight for all data points, and  
   the fitted lines were then translated in the y-direction to pass through (x, y)=(0.55, 1). 
   The 4.2\%\,$\upmu$m$^{-1}$ uncertainty allows distinguishing between some classes of weakly-featured asteroids ({\it left panel}). Other classes, such as the Xk and C-type, or the silicate-rich S- and Q-type asteroids, overlap in spectral slope ({\it right panel}): a combined slope+band analysis, or a Principal Component Analysis, would ideally be required for reliable classification of objects in these classes. The 4.2\%\,$\upmu$m$^{-1}$ uncertainty further closely matches the spectral diversity of most taxonomic classes, with the exception of the S and D types, implying compositional variations cannot be easily distinguished from intrinsic uncertainties.} 
\label{fig:taxons}
\end{figure}

We further note that the 4.2\%\,$\upmu$m$^{-1}$ uncertainty derived in this work closely matches the 1$\upsigma$ spread in spectral slope of some taxonomic classes of asteroids (Table\,\ref{tab:taxons} and Fig.\,\ref{fig:taxons}). 
Most spectral variability observed within these taxons could thus be explained by scattering in the measurements, without the need to invoke intrinsic compositional heterogeneities 
(note that several classes listed in Table\,\ref{tab:taxons}, such as the Sa and T, exhibit a very small spread in spectral slope of $\sim$1 or 2\%\,$\upmu$m$^{-1}$ that is due to the statistically low number of objects in these spectral types). 
Other classes of objects, like the D and S-type asteroids, exhibit a broad spread in spectral slopes indicative of true compositional diversity within these groups. 
V-type asteroids, which originated from one or several collisions on (4)~Vesta for the most part (e.g., \citealt{Carruba:2005, Russell:2012}) also show significant slope variability, which may be linked to the compositional heterogeneity of the differentiated Vesta. 
Varying band depth in classes of strongly featured asteroids (e.g., Q, S, V-types) also certainly affects our slope measurements (here, all data points in the spectra, including those located in regions of strong silicate absorption, were used to calculate the slope). 

In general, instrumental effects are unlikely to be the sole contributor to slope diversity within individual asteroid taxons. 
Physical processes such as regolith grain size and surface age, which directly relates to the amount of space weathering \citep{Hapke:2001} an atmosphereless body experienced since its most recent resurfacing event, are known to induce systematic change of the reflectance slope for various classes of asteroids and meteorites (e.g., \citealt{Nesvorny:2005, Strazzulla:2005, Brunetto:2006, Lazzarin:2006, Marchi:2006, Loeffler:2009,Vernazza:2009,Vernazza:2016,Cloutis:2011,Cloutis:2012,Fu:2012,Lantz:2013, Lantz:2017}). 
Instrumental effects, however, cannot be ignored if slope variation is not correlated to any physical parameter of the asteroids. 
More generally, uncertainties on slope measurements provided in this work should be accounted for in future classification works of small bodies and compositional investigation of individual objects to test whether observed slope variations reflect true compositional variations, or whether they can be attributed to instrumental effects. 

\begin{deluxetable}{l | cc | cc | c}
\caption{Average and 1$\upsigma$ spread of spectral slope measurements in the taxonomic classes of the \citet{demeo:2009} classification scheme, over the 0.45 to 2.45 (vnir) and 0.80 to 2.45 (nir) wavelength ranges. The spectra were normalized to 0.55\,$\upmu$m in the vnir, and 1.00\,$\upmu$m in the nir.
\label{tab:taxons}}
\tablehead{
\colhead{Spt. Type} & \colhead{${\rm s'_{vnir}}$} & \colhead{$\upsigma_{\rm s'_{vnir}}^*$} & \colhead{${\rm s'_{nir}}$} & \colhead{$\upsigma_{\rm s'_{nir}}^*$} & \colhead{$N^{\dagger}$} \\
\colhead{} & \colhead{[\%\,$\upmu$m$^{-1}$]} & \colhead{[\%\,$\upmu$m$^{-1}$]} & \colhead{[\%\,$\upmu$m$^{-1}$]} & \colhead{[\%\,$\upmu$m$^{-1}$]} & \colhead{}
}
\startdata
A & 78.4 & 12.2 & 84.2 & 12.5 & 6 \\
B & -9.9 &  6.2 & -8.9 & 6.5 & 4 \\
C & 12.3 &  6.3 & 14.9 & 10.8 & 13 \\
Cb & 15.0 &  5.8 & 16.3 &  6.9 & 3 \\
Cg & 11.5 & -- & 12.5 & -- & 1 \\
Cgh & 11.2 &  3.8 & 8.7 & 2.8 & 10 \\
Ch &  7.8 &  4.6 & 7.6 &  7.2 & 18 \\
D & 61.2 & 21.6 & 39.9 & 18.4 & 16 \\
K & 12.7 &  5.7 & 10.5 &  6.5 & 16 \\
L & 12.4 &  8.3 & 3.2 & 11.6 & 22 \\
O &  5.5 & -- & 18.4 & -- & 1 \\
Q & 13.6 &  6.1 & 21.1 & 10.7 & 8 \\
R & 31.5 & -- & 35.3 & -- & 1 \\
S & 20.7 &  8.6 & 14.2 & 10.0 & 144 \\
Sa & 43.7 &  1.3 & 53.6 &  1.7 & 2 \\
Sq & 21.5 &  9.5 & 21.0 & 14.4 & 29 \\
Sr & 20.4 &  9.6 & 15.8 & 12.1 & 22 \\
Sv & 22.8 &  4.7 & 11.1 & 10.2 & 2 \\
T & 30.4 &  2.3 & 20.0 &  1.9 & 4 \\
V & 13.4 & 13.8 & 10.0 &  9.1 & 17 \\
X & 27.3 &  2.5 & 22.1 &  0.6 & 4 \\
Xc & 11.1 &  6.2 & 7.3 & 13.1 & 3 \\
Xe & 11.8 &  4.0 & 6.8 &  2.1 & 7 \\
Xk & 19.3 &  7.1 & 15.5 &  9.4 & 18 \\
\enddata
 \tablecomments{${}^*$Provided only for taxonomic classes with at least two reported members in \citet{demeo:2009}. ${}^{\dagger}$Number of objects used to calculate the slopes (i.e., number of classified objects in the Bus-DeMeo taxonomic system).}
\vspace{2mm}
\end{deluxetable}

\section{Summary} \label{sec:summary}

We investigated accuracy limits of spectral slope measurements and systematic errors in the NIR (0.8--2.4\,$\upmu$m) SMASS and MITHNEOS asteroid surveys using 628 spectra of 11 solar analogs. 
Our main findings can be summarised as follows: 

\begin{itemize}
    \item The intrinsic 1$\upsigma$ slope uncertainty over 0.8--2.4~$\upmu{\rm m}$ is $4.2\%\,\upmu {\rm m}^{-1}$. 
    This value was derived after correcting the long-term variability of the telescope optical transmission function and the detector response function, 
    which is removed in asteroid surveys by the process of dividing the asteroid spectrum by a solar spectrum acquired on the same night through the same instrumental configuration. 
    \item Spectral slope decreases on average by 0.92\%\,$\upmu$m$^{-1}$ per 0.1 unit air mass difference between the asteroid and the solar analog star. 
    This effect accounts for only $\sim$2.8\%\,$\upmu$m$^{-1}$ variability in the SMASS and MITHNEOS surveys, considering that they are designed to operate mostly in the 1.0 to 1.3 air mass range.
    \item Slit alignment can be a major source of uncertainty in slope measurements of small bodies. This effect can be mitigated by acquiring multiple calibration stars allowing the identification of outlier measurements, and by re-aligning the target on the slit along the observations to monitor slope variability as a function of centering. 
    \item Additional possible sources of systematic uncertainties explored in this work, including weather conditions, parallactic angle, and the choice of the calibration star used to divide the asteroid spectra, were not found to contribute to systematic slope change in our dataset at the $>$2$\upsigma$ confidence level. 
    \item The overall 4.2\%\,$\upmu$m$^{-1}$ uncertainty on slope measurement 
    allows distinguishing between some classes of weakly-featured asteroids (B-, C-, X- and D-types), while others (e.g., C versus Xk) require an analysis of the overall shape of the continuum and of the weak features in the spectra, and/or a PCA similar to that performed in \citet{demeo:2009}. 
    The spread in spectral slope observed in several classes of asteroids closely matches the statistical uncertainty (Table~\ref{tab:taxons}), implying that these classes could be compositionally homogeneous 
    (this, of course, is not true for classes with varying spectral features). 
    Slope uncertainties derived in this work should be considered in future classification works and compositional investigation of small bodies to differentiate true intrinsic composition heterogeneities from possible instrumental effects. 
\end{itemize}

\section*{Acknowledgements}
Based on observations collected at the Infrared Telescope Facility, which is operated by the University of Hawaii under contract NNH14CK55B with the National Aeronautics and Space Administration. The authors acknowledge the sacred nature of Maunakea,
and appreciate the opportunity to observe from the mountain. Observations were conducted remotely mainly from the MIT-IRTF remote observing facility. M.M. and F.D. were supported by the National Aeronautics and Space Administration under Grant No. 80NSSC18K0849 issued through the Planetary Astronomy Program. 

\bibliographystyle{aasjournal.bst}
\bibliography{references} 

\appendix
\section{Measured 0.8--2.4-micron spectral slope \textit{versus} observing conditions}
\label{sec:app_A}

\begin{figure}[ht]
   \centering
   \includegraphics[width=0.45\linewidth]{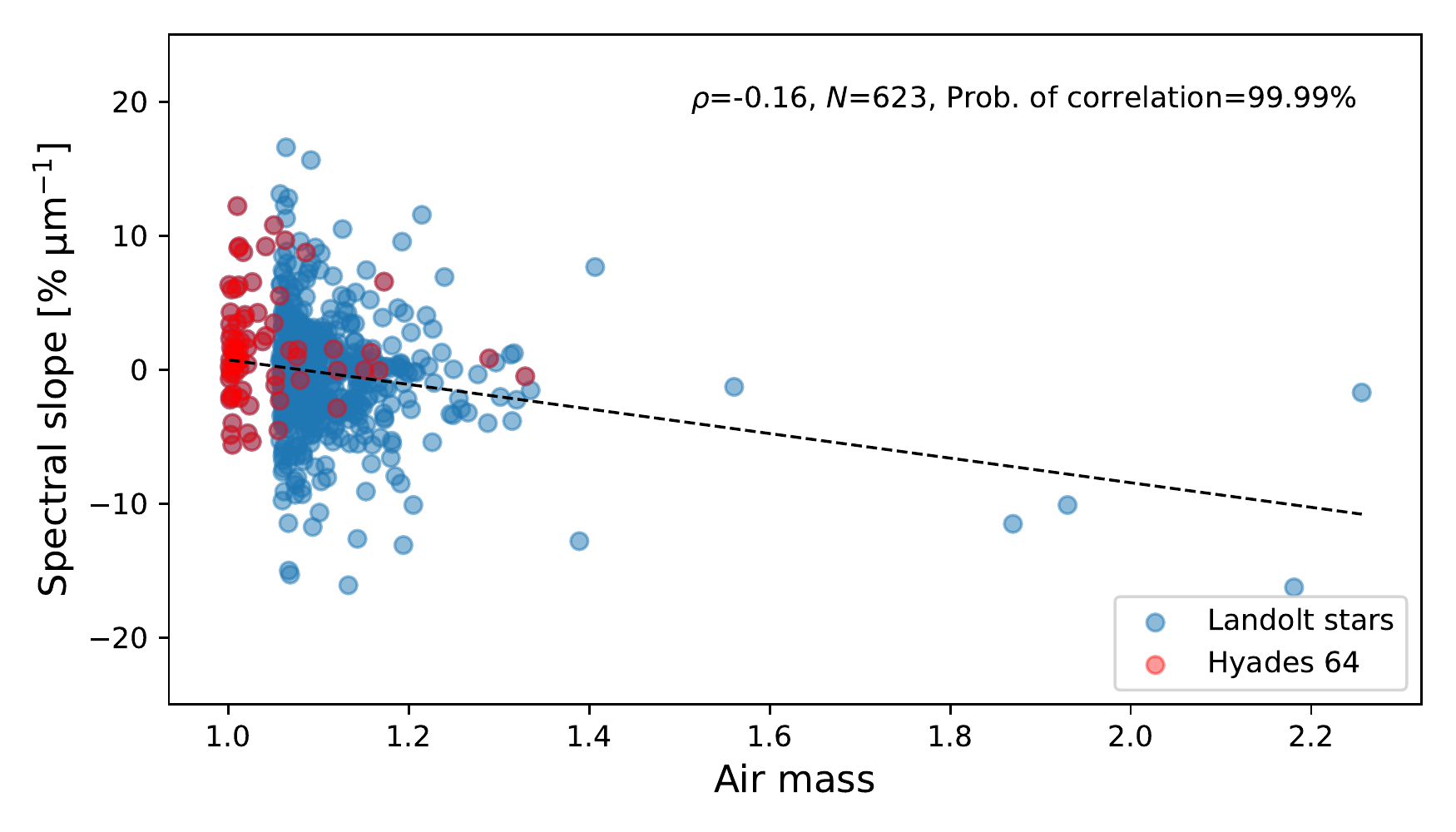}
   \includegraphics[width=0.45\linewidth]{sup_AMlow_slope.pdf}
   \includegraphics[width=0.45\linewidth]{sup_Parang_slope.pdf}
   \includegraphics[width=0.45\linewidth]{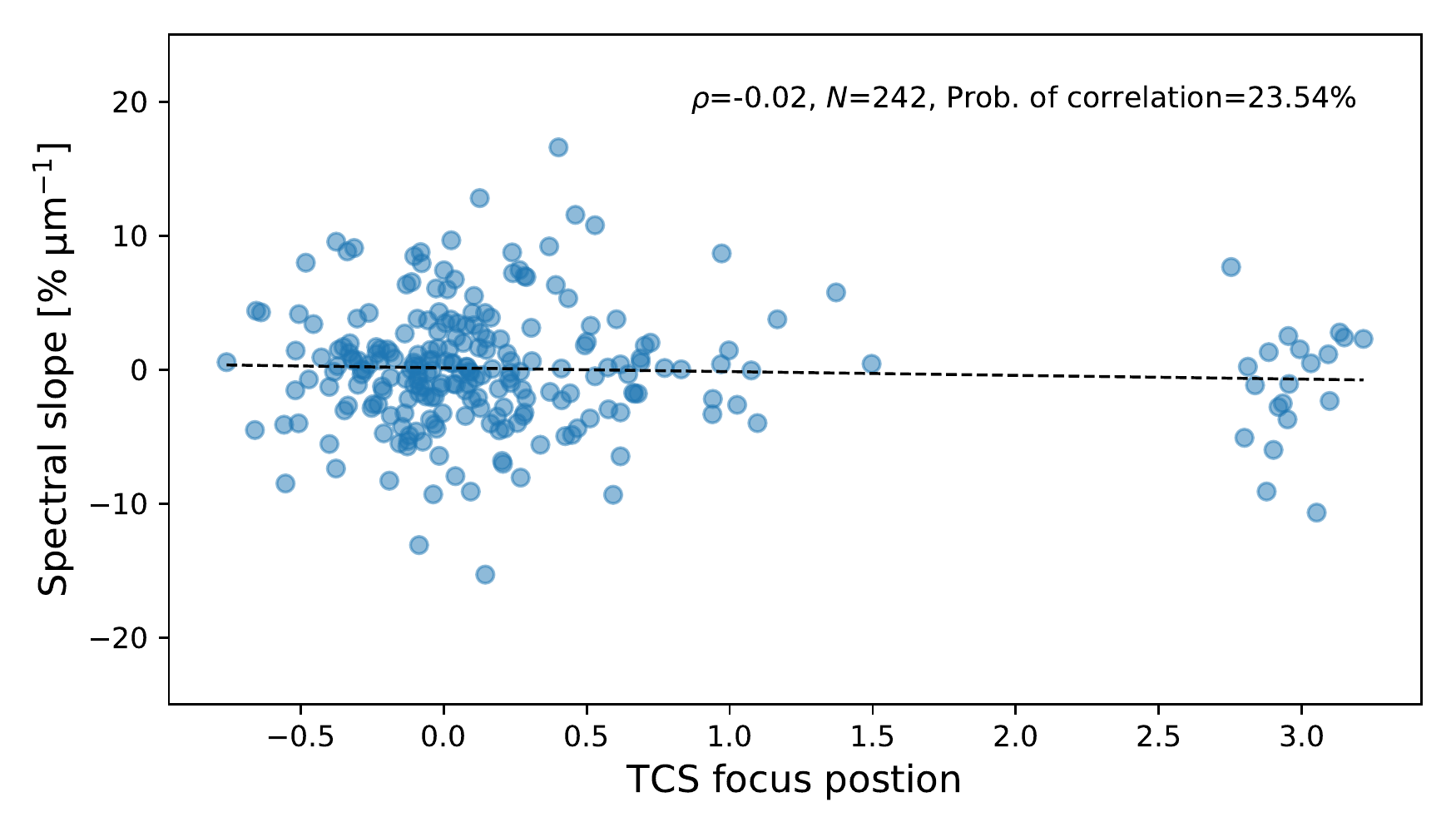}
   \includegraphics[width=0.45\linewidth]{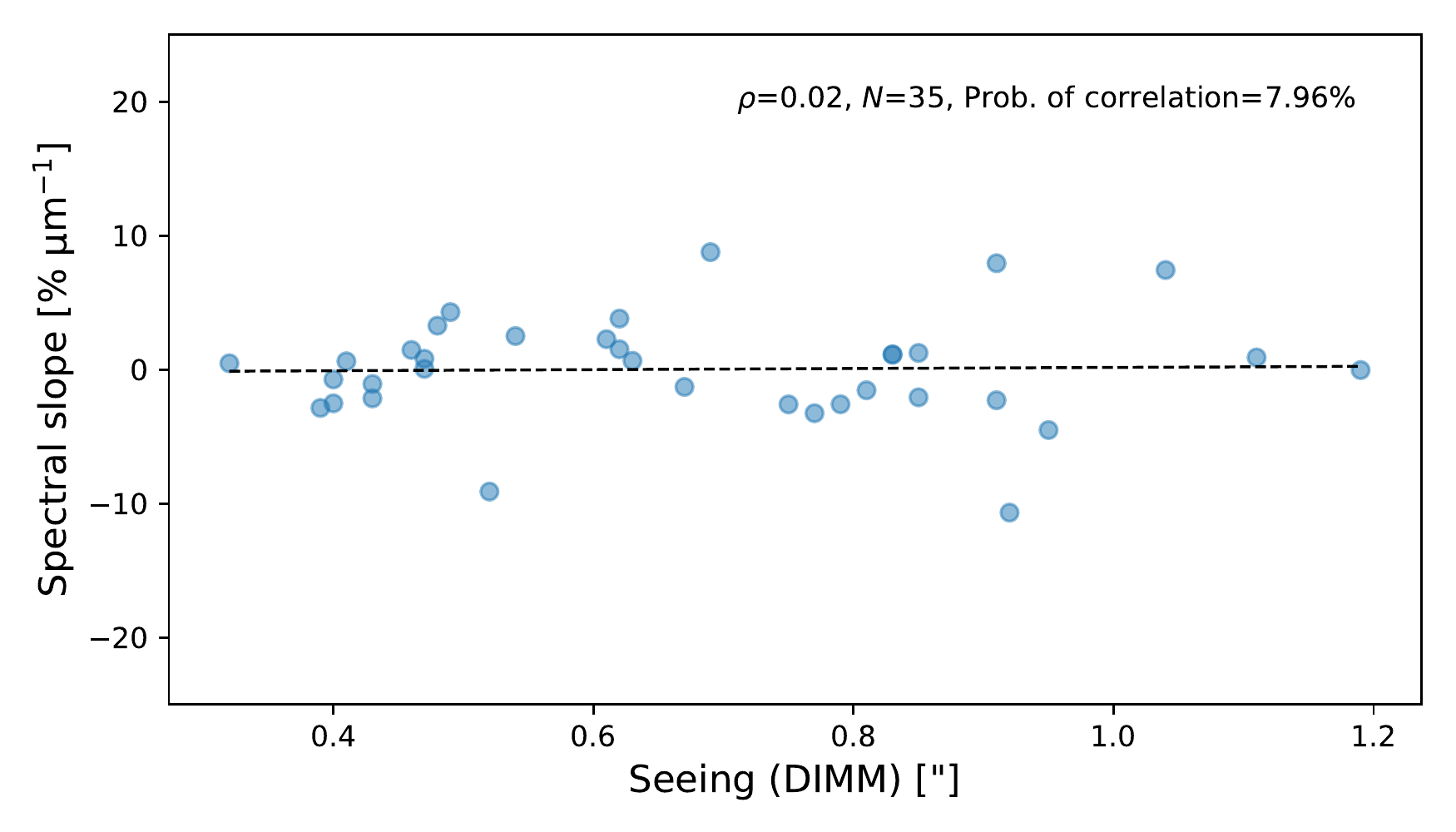}
   \includegraphics[width=0.45\linewidth]{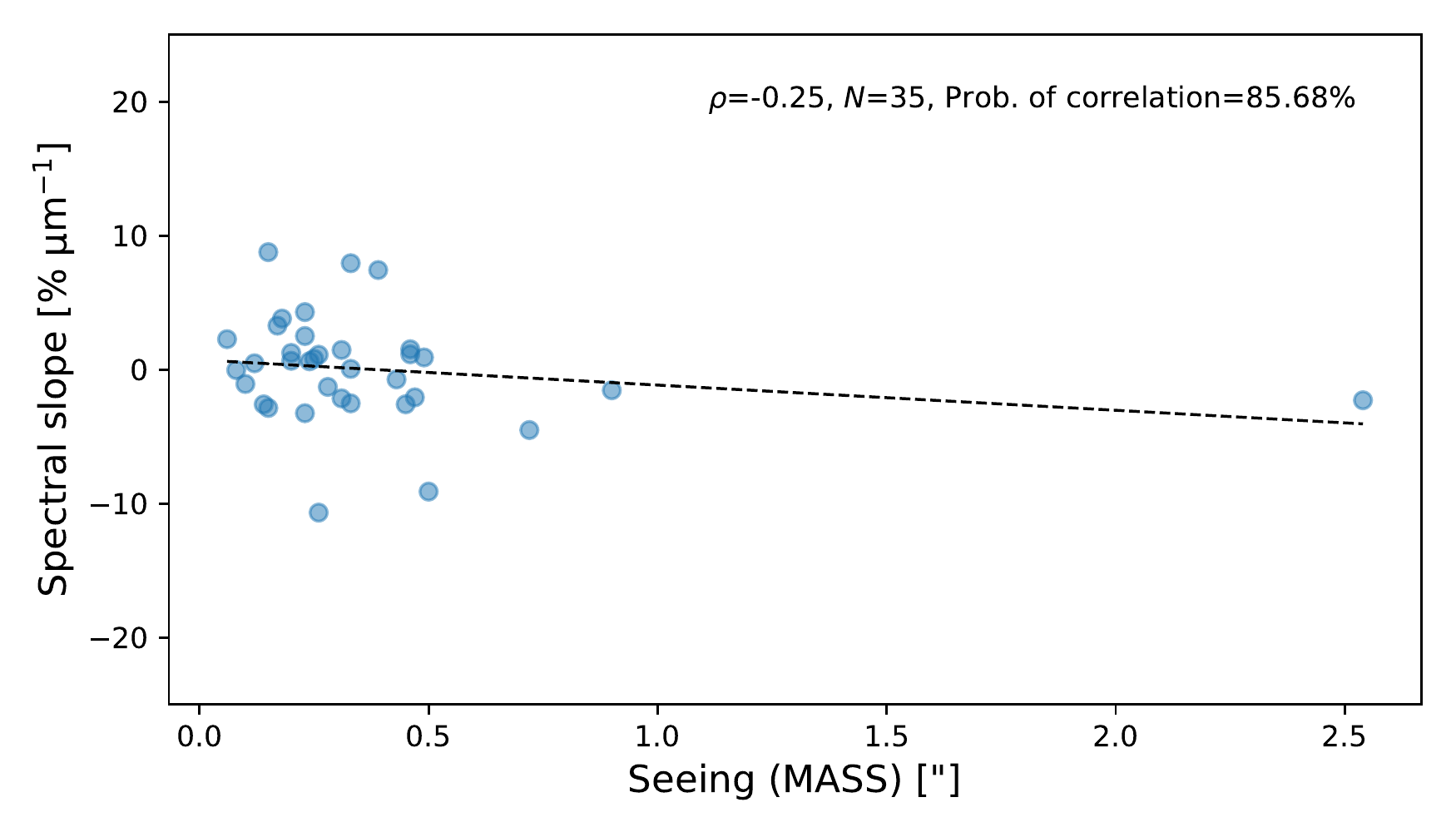}
   \caption{Measured 0.8--2.4-$\upmu$m spectral slope as function of observing conditions and instrument parameters. Each panel displays a distinct parameter, with the dotted line showing linear fit to the data. Air mass is shown twice: One panel includes all measurements, and the other is a zoom on the $\sim$1.0--1.3 air mass range where most measurements ($\sim$97.5\%) were collected. Some parameters (e.g., seeing measurements) have been recorded only for the most recent measurements. A description of the various parameters is available at \url{http://smass.mit.edu/minus.html}. The Spearman rank-order correlation coefficient ($\rho$), number of observations ($N$) and corresponding probability of correlation, computed as 1 minus the two-sided p-value, are indicated in each panel. The p-value depends both on $\rho$ and $N$: as sample size decreases (e.g., in the case of seeing measurements where only a few measurements are recorded), a higher correlation parameter is needed to reach a given probability. Air mass is the only parameter found to correlate with measurements at a significant statistical level, with -0.92\%\,$\upmu$m${}^{-1}$ slope variation per 0.1 unit air mass.} 
\label{fig:app_A1}
\end{figure}

\begin{figure}[ht]
   \ContinuedFloat
   \centering
   \includegraphics[width=0.45\linewidth]{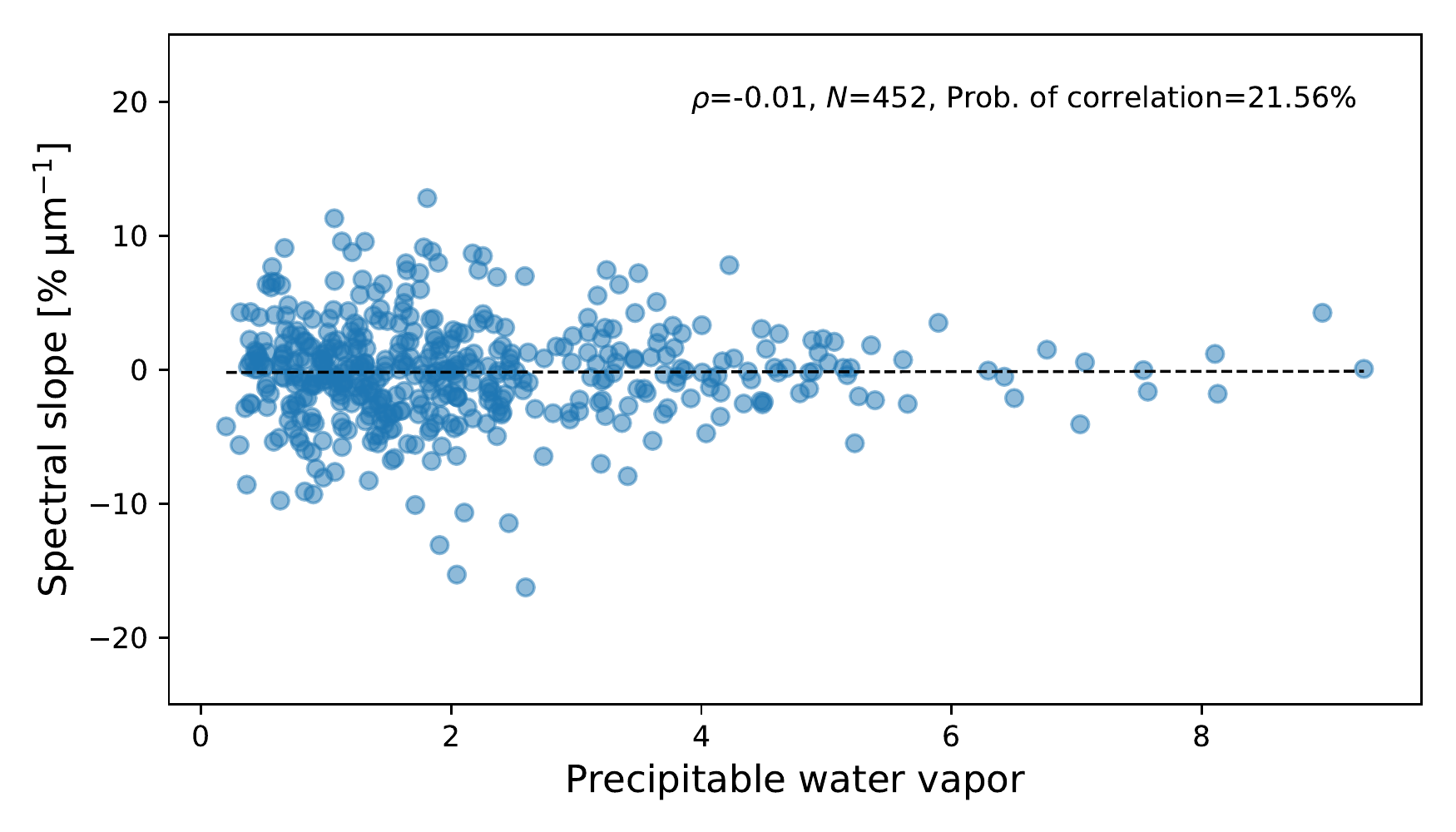}
   \includegraphics[width=0.45\linewidth]{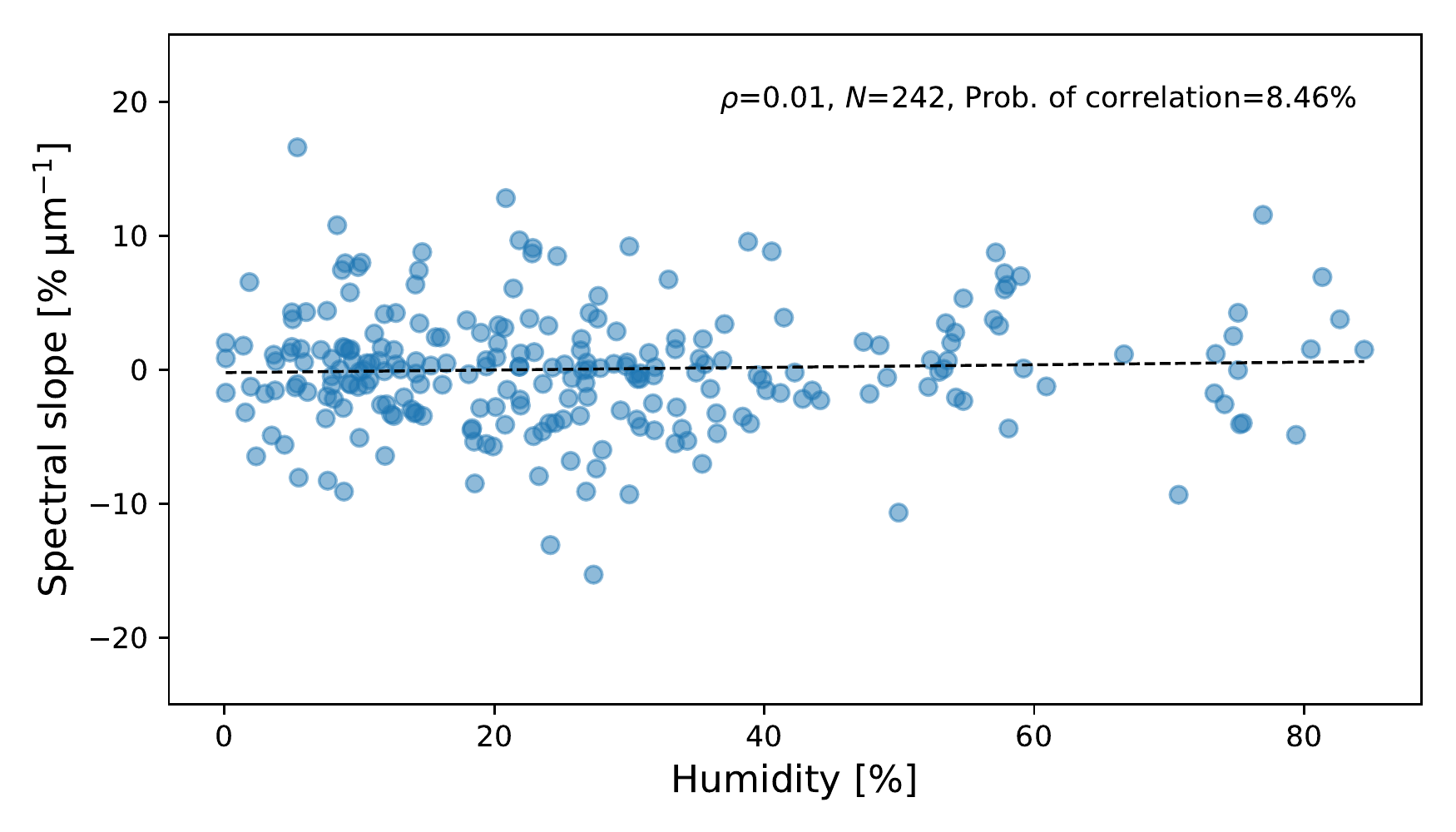}
   \includegraphics[width=0.45\linewidth]{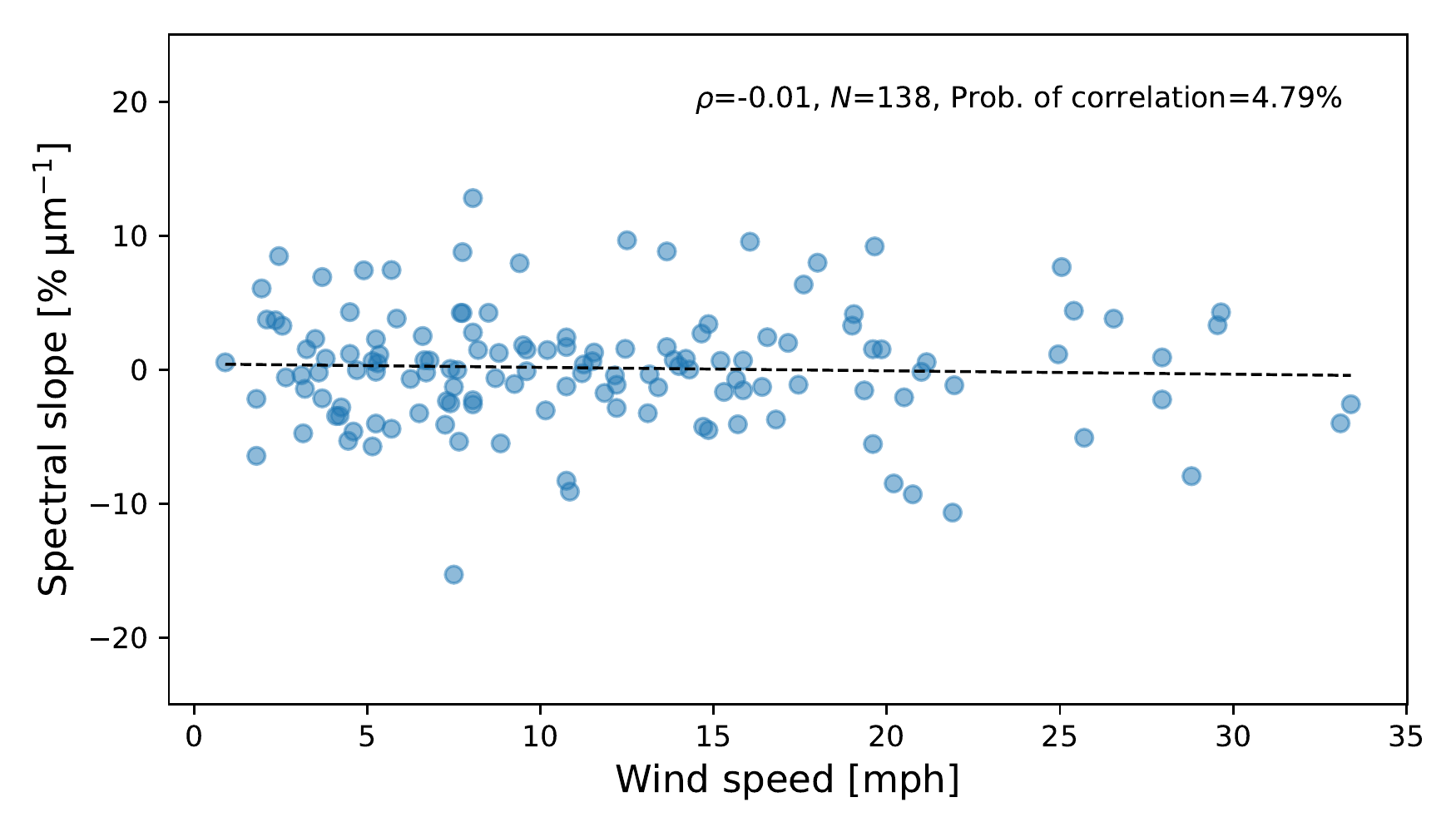}
   \includegraphics[width=0.45\linewidth]{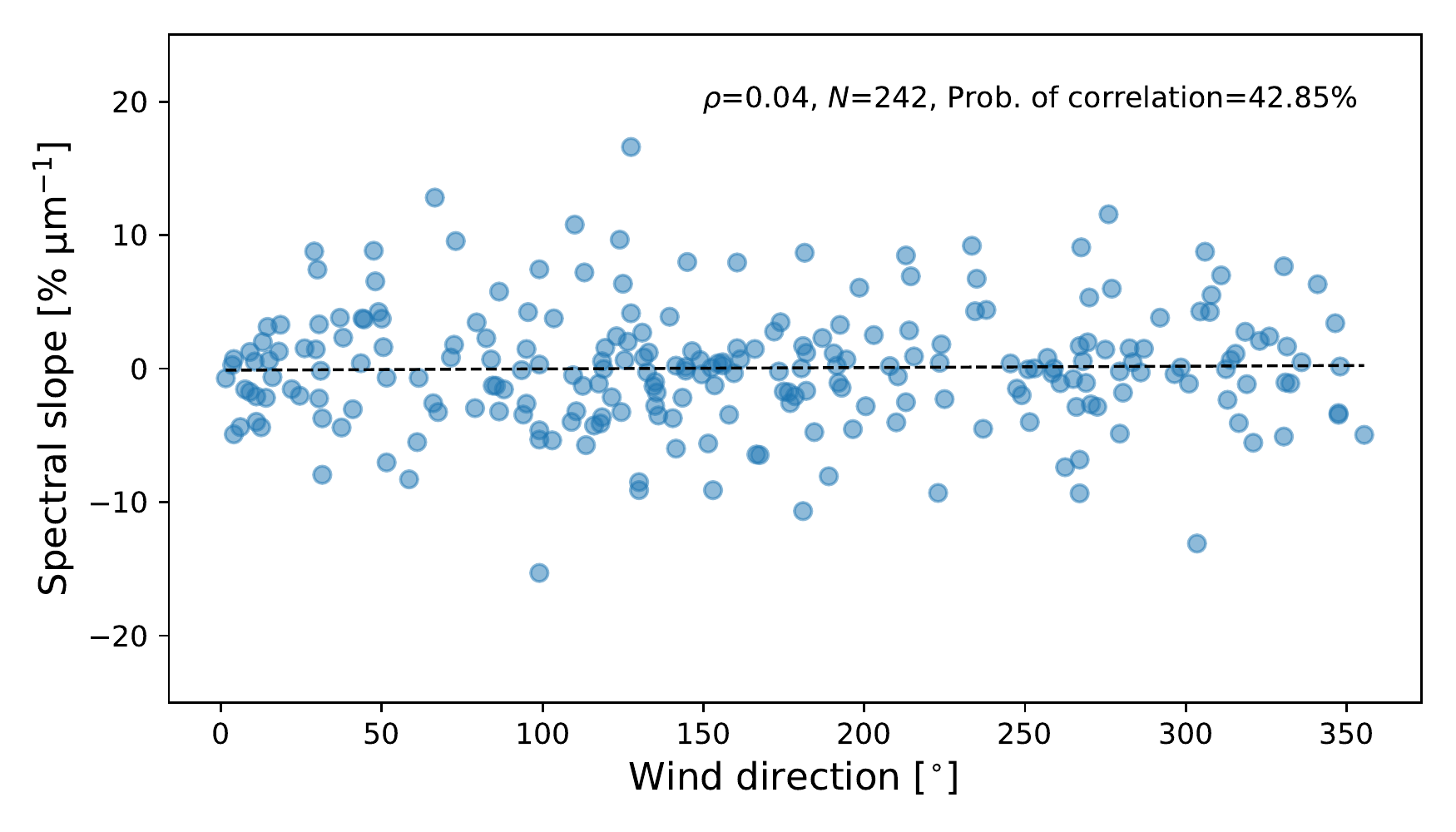}
   \includegraphics[width=0.45\linewidth]{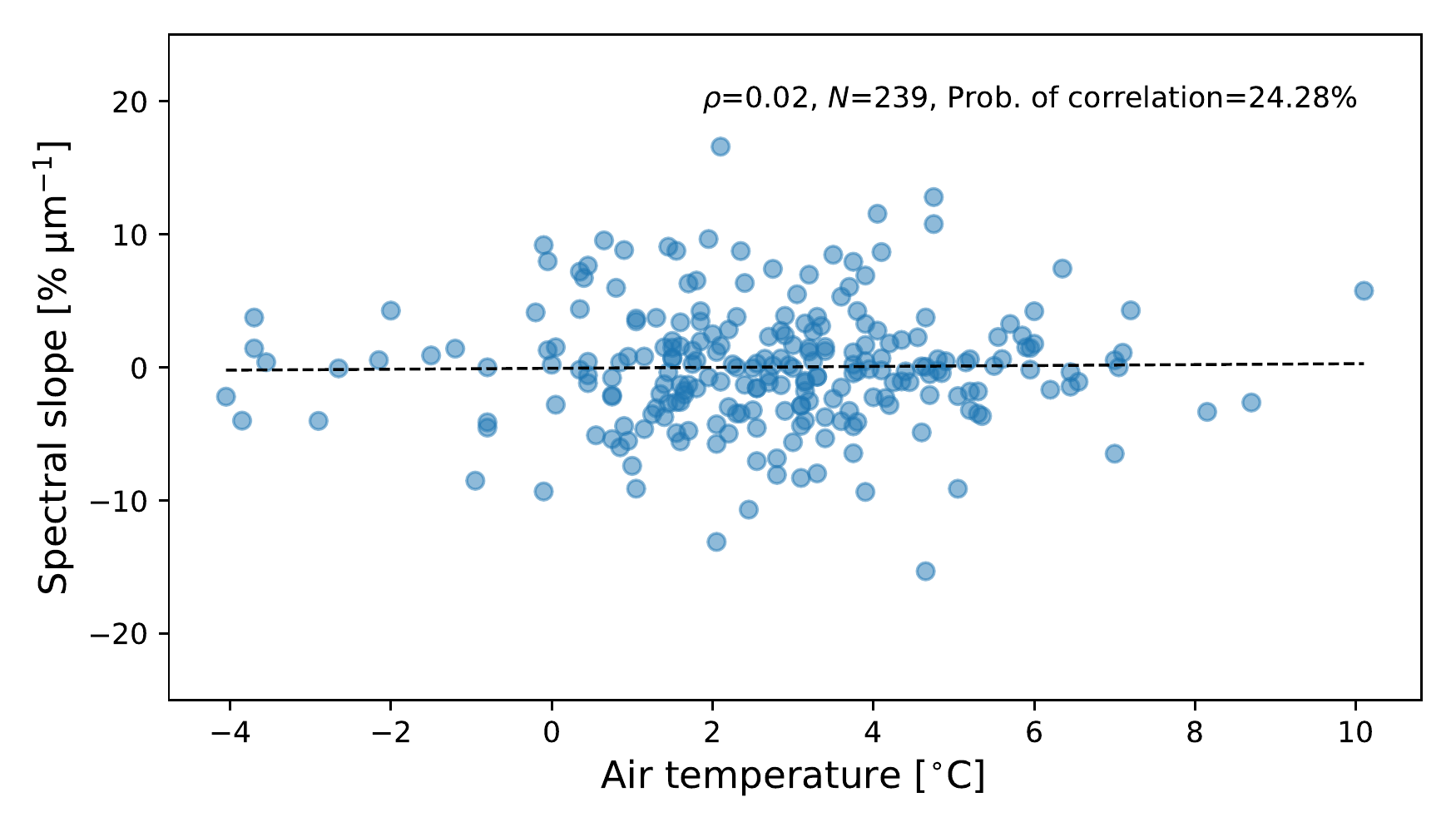}
   \caption{\textit{continued}.}
\label{fig:app_A2}
\end{figure}

\end{document}